# A Two-Step Test to Identify Zero-Inflated Biomarkers in Early-Phase Clinical Trials

Nan Miles Xi [1*], Lin Wang [2]

[1] Data and Statistical Sciences, AbbVie Inc., North Chicago, IL 60064, USA

[2] Department of Statistics, Purdue University, West Lafayette, IN 47907, USA

* Correspondence: nan.xi@abbvie.com

## Abstract

In early-phase clinical trials, a predictive biomarker may identify subgroups that benefit from an experimental therapy, even when the overall average treatment effect is negligible. Recently proposed nonparametric interaction tests such as the Average Kolmogorov-Smirnov Approach (AKSA) avoid prespecified biomarker cutting points and model assumptions, but their power degrades when the biomarker distribution is zero-inflated. We propose a two-step test that partitions the analysis into a spike test for biomarker-negative patients and a tail test for biomarker-positive patients, then combines the resulting p-values using Fisher's or Brown's method. This design isolates distinct sources of predictive effects, mitigates dilution, and preserves exact type I error control through permutation calibration. We derive theoretical properties showing that the proposed test retains nominal size and improves power over AKSA when predictive effects are concentrated in either the spike or tail subpopulation. Extensive simulations confirm robust type I error control under various zero-inflation rates, sample sizes, and skewed biomarker distributions. We also demonstrate consistent power gains across spike-only, tail-only, and mixed-effect scenarios. Our method provides a practical and flexible tool for early-phase trials with sparse biomarker distributions, enabling more reliable identification of predictive biomarkers to guide later-phase development.

# 1 Introduction

Precision medicine has placed a major emphasis on identifying patient subgroups that can benefit from new therapies [1,2]. In early-phase clinical trials, it is common that only a subset of patients responds, even if the overall average treatment effect is negligible [3,4]. Investigators devote substantial effort to biomarker discovery in these proof-of-concept trials, aiming to detect predictive biomarkers that signal which patients are more likely to respond [5]. Modern trial designs also stratify patient cohorts by biomarker status to prospectively evaluate treatment effects within subgroups [6–8]. However, in many early-phase trials, the predictive value of a biomarker is uncertain upfront, and one often assesses biomarker-treatment interactions in retrospect. This creates a need for flexible statistical methods to evaluate potential biomarkers when the global null hypothesis of no treatment benefit may hold in the population.

Traditional interaction tests and subgroup analyses face challenges in this setting. They may require a prespecified biomarker cutoff or make parametric assumptions that are difficult to justify [9]. Recently, nonparametric approaches have been proposed to avoid these limitations. For example, the Average Kolmogorov-Smirnov Approach (AKSA) was developed as an interaction test that does not require choosing a single cutting point [10]. AKSA uses label permutation to obtain exact p-values under the global null and compares treatment and control outcomes across the entire range of biomarker values. This data-driven method is well suited for detecting a broad class of treatment-biomarker interactions without model assumptions [11]. In practice, though, AKSA and similar methods encounter limitations when the biomarker distribution is zero-inflated [12].

Zero-inflation is a feature that the biomarker is undetectable in a large fraction of subjects. It can dilute the signal that a predictive biomarker test is trying to capture [13]. When a large subset of trial participants has a biomarker value of zero, the information that AKSA relies on becomes heavily concentrated at that single point. As a result, the average treatment-outcome signal extracted is diluted while the random noise contributed by the biomarker-null patients remains. This phenomenon leads to loss of power, making the test have a harder task, and its sensitivity erodes as the proportion of zero values increases. Since the original AKSA gives equal weight to every portion of the biomarker range, the large spike at zero reduces the overall contrast and yields a weaker signal than one would get by examining the zero and non-zero parts separately. This signal dilution challenge under strong sparsity calls for a tailored solution.

In this study, we propose a two-step test that explicitly addresses the spike-versus-tail issue in zero-inflated biomarker data. In our approach, the predictive biomarker hypothesis is partitioned into two complementary components. First, a spike test assesses the treatment effect in the subgroup of patients whose biomarker equals zero. This component asks whether outcomes differ between treatment and control among biomarker-negative individuals. Second, a tail test focuses on patients with biomarker values above zero, i.e., the positive tail of the distribution. The tail test applies a modified AKSA on this subset, measuring treatment-control differences across increasing biomarker levels. Each of the two components produces a p-value through exact permutation inference under the global null which assumes no treatment effect in any subgroup. We then combine the evidence from the spike and tail through a suitable combination rule: Fisher's combination is used when the two tests are independent, while Brown's combination provides an adjustment for positive correlation between component tests [14,15]. This two-step combination generates a single overall test for the presence of a predictive biomarker effect, while still allowing the spike and tail evidence to contribute separately.

The proposed two-step test resolves the issue of a diluted signal by analyzing both the zero spike and the continuous tail of the biomarker. One benefit of our proposed test is the power gain. If a treatment effect is concentrated in either the zero-inflated or biomarker-positive subpopulation, the two-step test can capture it effectively, whereas a one-part method would risk missing it by averaging the effects. This approach retains the inferential properties of permutation testing. By shuffling treatment labels to calibrate significance, we achieve the control of type I error at the nominal level without large-sample approximations or parametric distributional assumptions. The method also remains robust to skewed biomarker distributions. Notably, the two-step test has proven effective in other fields with sparse data. For example, zero-inflated models that separately test the occurrence and magnitude of an outcome have shown improved power and robustness in microbiome and genomic analyses [16,17]. Building on this insight, our two-step test provides a novel tool for biomarker-driven clinical trials, ensuring that a large proportion of zero readings does not mask a potentially actionable treatment effect in a subset of patients.

The remainder of this paper is organized as follows: Section 2 reviews AKSA, formalizes the zero-inflated biomarker setting, and states the working assumptions. Section 3 introduces the proposed two-step test, describes Fisher's and Brown's combination, and proposes post-rejection diagnosis and cut-point selection. Section 4 develops theoretical properties of exact type-I error control and power gain. Section 5 reports a comprehensive simulation study, demonstrating type I error control under null, power under various alternatives, test dependence, and robustness to skewed biomarkers. Section 6 offers a brief discussion.

## 2 Limitation of AKSA in Zero-Inflated Settings

In proof-of-concept trials, it is often unknown whether an experimental treatment confers any average benefit. The global null therefore assumes no treatment effect in the full population. Under this sharp null, AKSA employs label permutations to compute exact p-values [10,18]. When a biomarker takes the zero value in a large fraction of subjects, the information that AKSA relies on becomes heavily compressed at this single point [19]. Because the method measures treatment-outcome differences by repeatedly comparing subjects across the entire range of biomarker values, a large concentration at zero leaves fewer informative contrasts [12]. As a result, the average signal that AKSA extracts is diluted, but the random variation contributed by the many zero subjects remains [20]. The test has a harder task and its detection power erodes even if the biomarker is genuinely predictive among the non-zero subjects. Moreover, the usual remedy of increasing the sample size provides only partial relief [21]. To see this, let the biomarker distribution be

$$X \sim \pi_0 \delta_0 + (1 - \pi_0) H \quad \text{with} \quad 0 \leq \pi_0 < 1$$

where $\delta_0$ is a point mass at zero, $\pi_0$ is the zero-inflation rate, and $H$ is a distribution on $(0, +\infty)$. Define the treatment effect curve

$$D(x) = \text{logit} \Pr(Y = 1 \mid T = 1, X = x) - \text{logit} \Pr(Y = 1 \mid T = 0, X = x)$$

and let $\Delta(\pi_0) = E_X(D(X))$ be its population average. Assume

**A1.** $D(x)$ is nondecreasing in $x$ and satisfies $D(x) > D(0)$ for some $x > 0$.

**A2.** Under global null, the AKSA statistic obtained by permuting treatment labels attains level $\alpha$ and its first-order variance is constant with respect to $\pi_0$.

Assumption **A1** formalizes a biologically plausible ordering: if the biomarker is truly predictive, larger positive values typically amplify the treatment benefit measured on the log-odds scale. Assumption **A2** is based on well-established properties of permutation calibration: under the global null, permuting treatment labels renders the empirical distribution of the AKSA statistic free of the unknown outcome model and driven only by sample size to first order [11]. Under such conditions, we formalize the power loss of AKSA under zero inflation in the following theorem:

> **Theorem 1. AKSA power loss under zero-inflated biomarkers**
>
> Denote by $\beta_{\text{AKSA}}(\pi_0)$ the power of the AKSA test with sample size $N$. Then $\beta_{\text{AKSA}}(\pi_0)$ are strictly decreasing functions of $\pi_0$. That is, $\beta_{\text{AKSA}}(\pi_0) < \beta_{\text{AKSA}}(0)$ for every $\pi_0 > 0$.

The complete proof of **Theorem 1** is provided in the **Appendix**. Intuitively, AKSA's power depends on the average treatment effect $\Delta(\pi_0)$. Introducing zero inflation moves probability mass to $X = 0$, the region where the treatment effect $D(x)$ is smallest under assumption **A1**. This shift lowers $\Delta(\pi_0)$ and the noncentrality parameter of the underlying $Z$-statistic shrinks while the variance term stays nearly constant by assumption **A2**. Because the rejection threshold is fixed, the test's power falls monotonically with each additional unit of zero inflation and ultimately approaches the nominal size $\alpha$ when almost all observations have $X = 0$.

To evaluate how zero inflation influences the ability of AKSA to detect a predictive biomarker, we perform a fully crossed Monte Carlo experiment. For each combination of zero-inflation rate $\pi_0 \in \{0, 0.1, \ldots, 0.8\}$, total sample size $N \in \{60, 90, 120\}$, and effect magnitude $\Delta \in \{3, 4, 5\}$. In every dataset, $N\pi_0$ subjects have biomarker $X$ at 0, while the remaining $N(1 - \pi_0)$ subjects draw $X$ values from a uniform distribution $U(0,1)$. We subsequently permute the combined biomarker vector. Treatment is assigned at random with equal allocation and the outcome variable is sampled from a standard normal distribution $N(0, 1)$. To simulate a predictive biomarker, we induce a strictly monotone treatment effect by incrementing the outcomes for treated subjects with X > 0 by $\delta \times \frac{\text{rank}(X)}{N}$. Each replicate dataset is analyzed with AKSA, in which the observed statistic is compared with the reference distribution obtained from 1,000 random permutations of the treatment labels, and the resulting permutation p-value is recorded. We estimate the empirical power using the proportion of p-values less than 0.05 across the 1,000 replicates.

**Figure 1A** presents the empirical power of AKSA as a function of the zero-inflation rate $\pi_0$. Across all scenarios, power declines monotonically as $\pi_0$ increases from 0 to 0.8, confirming the theoretical result that concentrating mass at $X = 0$ attenuates the average treatment contrast and hence the test's sensitivity. At any fixed $\pi_0$, larger sample sizes yield higher power, while stronger interaction effects shift the entire power curve upward. Nevertheless, the downward slopes have a similar shape, indicating that neither increasing sample size nor enlarging effect fully counteracts the loss of power induced by zero inflation. The horizontal reference line at 0.8 highlights that, even under the most favorable setting ($\Delta = 5, N = 120$), power falls below the conventional 80% benchmark once $\pi_0$ exceeds approximately 0.35, and all curves converge toward the nominal level 0.05 as $\pi_0$ approaches 0.8.

## 3 Two-Step Test for Detecting a Predictive and Zero-Inflated Biomarker

We propose a two-step test for detecting a predictive biomarker when its distribution is zero-inflated and no average treatment benefit is present under the null. The test partitions the predictive biomarker question into (i) a *spike* test that evaluates the treatment contrast when the biomarker equals zero, and (ii) a *tail* test that applies AKSA to the strictly positive part of the biomarker distribution. The two resulting p-values are then synthesized by either Fisher's combination, which is valid under independence of two component tests, or Brown's combination, which remains accurate under positive correlation.

### 3.1 Spike test

Consider an i.i.d. sample

$$\{(Y_i, T_i, X_i): i = 1, \dots, N\}, \quad T_i \in \{0,1\}, \quad Y_i \in R, \quad X_i \geq 0$$

where $T_i$ indicates treatment, $Y_i$ is the continuous outcome, and the biomarker $X_i$ is continuous and subject to zero inflation. Denote

$$I_0 = \{i: X_i = 0\}, \quad I_+ = \{i: X_i > 0\}, \quad n_0 = |I_0|, \quad n_+ = |I_+|$$

The statistic of spike test is the absolute mean difference

$$A_N = |\bar{Y}_{10} - \bar{Y}_{00}|$$

with

$$\bar{Y}_{t0} = \frac{1}{n_{t0}} \sum_{i \in I_0, T_i = t} Y_i, \qquad n_{t0} = |\{i \in I_0 : T_i = t\}|$$

The null distribution of $A_N$ is generated by randomly permuting the four-level group labels $G_i = T_i + 2\mathbf{1}\{X_i > 0\}$, conditional on the observed $\{Y_i\}_{i=1}^N$. The resulting permutation p-value is

$$p_A = \Pr_{\text{perm}} \{A_N^* \geq A_N\}$$

where $A_N^*$ denotes the permuted test statistic. If every subject has $X_i > 0$, we set $p_A = 1$.

### 3.2 Tail test

This is the regular AKSA on the positive tail of the biomarker. We order the positive subset $I_+$ by biomarker magnitude $X_{(1)} \leq \cdots \leq X_{(n+)}$. For each prefix $k = 1, \ldots, n_+ - 1$, compute the two-sample Kolmogorov-Smirnov distance

$$D_k = \text{KS}(\{Y_{(j)} : j \leq k, T_{(j)} = 1\}, \{Y_{(j)} : j \leq k, T_{(j)} = 0\})$$

We set $D_k = 0$ when all included subjects come from the same treatment arm. The statistic of the tail test is the average of all $D_k$

$$B_N = \frac{1}{n_+ - 1} \sum_{k=1}^{n_+ - 1} D_k$$

Its p-value $p_B$ is obtained from the permutation of the treatment labels inside $I_+$

$$p_B = \Pr_{\text{perm}} \{B_N^* \geq B_N\}$$

where $B_N^*$ denotes the permuted test statistic. With fewer than five positive observations in $I_+$ we set $p_B = 1$.

### 3.3 Combining two component p-values

The two components of the proposed test target distinct but complementary features of a zero-inflated biomarker. Each test produces a valid permutation p-value that reflects only partial evidence against the global null. We design two strategies to aggregate these pieces of evidence to report a single inferential statement at the study level.

**(i) Fisher's combination.** When the spike and tail tests are independent under the null, the sum of the logarithms of their p-values follows a chi-square distribution with degrees of freedom twice the number of components [22]. It creates an exact combined test that controls nominal type I error while maintaining power whenever either part is sensitive to the alternative. The test statistic and p-value of the Fisher combination are constructed as

$$S_{\text{Fisher}} = -2(\log p_A + \log p_B), \qquad p_{\text{Fisher}} = \Pr\{\chi_4^2 \geq S_{\text{Fisher}}\}$$

**(ii) Brown's combination**. In practice, the spike and tail tests may share similar outcomes and treatment assignments, thus exhibiting positive correlation. Brown's combination adjusts the Fisher chi-square distribution by an estimated scale factor and corresponding degrees of freedom, explicitly accounting for this dependence. Denote $\rho$ as the Spearman correlation between $p_A$ and $p_B$, the scaling factor $c = 1 + \rho$, and degree of freedom $\nu = 4/c$. The test statistic and p-value of Brown's combination are constructed as

$$S_{\text{Brown}} = \frac{S_{\text{Fisher}}}{c}, \qquad p_{\text{Brown}} = \Pr\{\chi_\nu^2 \geq S_{\text{Brown}}\}$$

The resulting test remains exact when the correlation estimate is zero and becomes conservative otherwise. Therefore, it safeguards the overall type I error while remining the benefits of joint evidence aggregation. In practice, $\rho$ can be calculated using the same permutation replicates that are already generated to obtain each component p-value [23].

## 3.4 Post-rejection diagnosis

To distinguish whether rejection of the global null is driven by a uniform main treatment effect or by a genuine treatment-by-biomarker interaction, we recommend a short post-hoc diagnostic routine that preserves the exact type I error of the primary analysis while providing substantive insight. First, report the component p-values $p_A$ and $p_B$ alongside the combined p-value. Comparable evidence in the two components is consistent with a homogeneous effect, whereas an imbalanced pair signals effect modification. Second, test the marginal main effect $H_0: \mathrm{E}(Y|T=1) = \mathrm{E}(Y|T=0)$ with an independent two-sample permutation test. A significant result with $p_A \approx p_B$ supports a non-predictive, purely prognostic biomarker. Third, centre the outcomes by subtracting the estimated overall treatment effect and reapply the spike and tail test. The resulting "interaction-only" p-value isolates predictive information, inflating to insignificance if the earlier rejection is driven solely by the main effect. Finally, plot a nonparametric estimate of the treatment effect curve with a bootstrap confidence band as an exploratory visual check. A flat band corroborates a prognostic biomarker, whereas a monotone or varying profile confirms predictiveness. These complementary steps require only elementary resampling tools, add negligible computational cost, and ensure that scientific conclusions drawn from the two-step test remain transparent and well substantiated.

## 3.5 Cutting point selection

To translate the continuous biomarker into a binary decision rule, we select the threshold $\hat{\tau}$ that maximizes the estimated treatment effect difference between biomarker strata. For every candidate split $\tau$, we compute the stratum-specific treatment contrasts, denoted as $\widehat{\Delta}_{\leq\tau}$ for $X \leq \tau$ and $\widehat{\Delta}_{>\tau}$ for $X \geq \tau$. The objective function

$$\hat{C}(\tau) = \left|\widehat{\Delta}_{>\tau} - \widehat{\Delta}_{\leq\tau}\right|$$

is evaluated over all unique biomarker values that leave at least five treated and five control subjects in each stratum. We set the cutting point as

$$\hat{\tau} = \underset{\tau}{\operatorname{argmax}}\, \hat{C}(\tau),$$

and then use the same permutation replicates generated for the two-step test to obtain an exact null distribution for $\hat{C}(\tau)$ and to construct bootstrap confidence interval for both $\hat{\tau}$ and the associated effect estimates. This approach aligns directly with the clinical objective of maximizing differential benefit, respects the zero-inflated structure by allowing the optimum to fall at the spike boundary or within the positive tail, and inherits valid finite-sample inference from the permutation framework [24].

## 4 Statistical Properties

In this section, we demonstrate the key features that make the proposed two-step test a reliable tool for detecting predictive biomarkers in early-phase trials where no overall treatment benefit is assumed a priori. First, Fisher's combination yields a single decision rule that retains exact type I error control when the two tests are independent. Second, Brown's combination remains slightly conservative type I error control when the two tests exhibit positive correlation. Third, under alternatives in which the treatment difference is concentrated either at $X = 0$, in the positive tail, or in both regions, the two-step test outperforms the original AKSA. It can reject whenever either component test captures signals, while AKSA relies solely on the positive tail and therefore loses power as zero inflation grows.

### 4.1 Control of type I error for Fisher's combination

Under the global null hypothesis, the two component tests are mutually independent and each yields a p-value that exactly follows a uniform distribution $U(0,1)$ [18]. Fisher's combination aggregates these into the statistic $S_{\text{Fisher}} = -2(\log p_A + \log p_B)$. Because the sum of two independent $-2\log p$ random variables follows a chi-square distribution with four degrees of freedom, the reference distribution of $S_{\text{Fisher}}$ is known in closed form and does not rely on large-sample approximations [25]. Consequently, rejecting whenever $S_{\text{Fisher}}$ exceeds the $(1-\alpha)$ quantile of $\chi_4^2$ guarantees that the combined test inherits exact size $\alpha$ regardless of sample size, zero-inflation rate, or other parameters. This property is formalized below:

> **Theorem 2. Fisher exact type I error control**
>
> Let $p_A$ and $p_B$ be the p-values from the *independent* spike and tail tests, respectively. Under $H_0$, $p_A, p_B \sim U(0,1)$. Define the rejection region $R_\alpha = \{S_{\text{Fisher}} \geq \chi^2_{4,1-\alpha}\}$. Then for every $0 < \alpha < 1$
>
> $$\Pr_{H_0}\{S_{\text{Fisher}} \in R_\alpha\} = \alpha$$
>
> That is, the two-step test that employs Fisher's combination is exactly level $\alpha$.

The complete proof of **Theorem 2** is provided in the **Appendix**. It shows that when the spike and tail tests are run on disjoint subsets, their p-values behave like two independent uniform draws under the global null, and Fisher's combination preserves this exact uniformity. In practice, this means investigators can combine evidence from the zero spike and the positive tail without worrying about an inflated false-positive rate. Whatever significance level they choose, the overall test adheres to it regardless of sample size, biomarker prevalence, or effect heterogeneity. The only limitation is the independence assumption, which breaks if the spike and tail tests are conducted from overlapping or correlated data. In such situations, the two component p-values are positively correlated, and Fisher's combination can become slightly anti-conservative [14]. In **Section 4.2** we therefore turn to Brown's combination, which preserves nominal size even under positive dependence.

### 4.2 Control of type I error for Brown's combination

While Fisher's combination offers an exact p-value when the two component tests are independent, empirical studies sometimes reveal a positive correlation $\rho$ between $p_A$ and $p_B$ [23,26]. To safeguard type I error in such circumstances, Brown's combination scales the Fisher statistic by a factor $c = 1 + \rho$ and replaces the $\chi_4^2$ reference distribution with a $\chi^2_{4/c}$ distribution obtained by moment matching [23]. When the correlation $\rho$ is zero, Brown's combination reduces exactly to Fisher. When $\rho > 0$, the scaling inflates the variance of the reference distribution, rendering the combined test slightly conservative rather than liberal. The next theorem formalizes these properties and shows that Brown's combination preserves nominal type I error control.

> **Theorem 3. Brown conservative type I error control**
>
> Let $p_A$ and $p_B$ be the p-values from the *positively correlated* spike and tail tests, respectively, with correlation $\rho > 0$. Under $H_0$, $p_A, p_B \sim U(0,1)$. Define the rejection region $R_\alpha = \{S_{\text{Brown}} \geq \chi^2_{v,1-\alpha}\}$, where $v = 4/c$ and $c = 1 + \rho$. Then for every $0 < \alpha < 1$
>
> $$\Pr_{H_0}\{S_{\text{Brown}} \in R_\alpha\} \leq \alpha$$

> That is, the two-step test that employs Brown's combination is conservative for any positve correlation between the component p-values.

The complete proof of **Theorem 3** is provided in **Appendix**. Essentially, Brown's combination complements Fisher's exact test by providing a built-in safety margin whenever the spike and tail components share information. The procedure preserves nominal size under independence and becomes mildly conservative when the correlation is positive. Simulation results in **Section 5** confirm that the associated loss of power is modest, whereas the protection against type I inflation is substantial. In practice, users can report both combined p-values, using the Fisher value as the primary decision metric, and present the Brown value as a sensitivity check that reassures readers of robustness to any dependence between the component tests. It is also beneficial to default to Brown's combination whenever independence is uncertain, especially in small samples or exploratory settings, confident that the overall type I error will stay at or below the nominal level without a costly sacrifice of power.

### 4.3 The power dominance of the two-step test

A zero-inflated biomarker can express predictive information in two qualitatively different places. Spike signal refers to a treatment contrast that appears exactly at the mass-point $X = 0$. Consistent with the notation used earlier, this is the value $D(0)$. Tail signal, by contrast, captures the pattern that the treatment effect rises progressively with the positive biomarker values, that is, $D(x)$ for $x > 0$. Either source may be present or absent, giving three practically relevant alternatives: spike-only, tail-only, and mix. This simple taxonomy underlies the design of our two-step test: spike test is tailored to detect spike contrast, tail test to detect tail contrast, and the Fisher/Brown combinations aggregate evidence so that power is gained whenever one or both signals are truly present.

Raw AKSA orders all subjects by increasing biomarker values and compares treated versus control outcomes for every prefix of the ordered list. It then averages those Kolmogorov-Smirnov distances over the entire set of prefixes. In the early prefixes where $X = 0$ for the first few subjects, the spike contrast is isolated, but these prefixes are numerically small, so their weights in the final average is limited. As soon as positive $X$ values enter the ordering, each subsequent prefix contains a mixture of spike and tail subjects. The pure spike signal is therefore blended with data from the tail and attenuated roughly in proportion to zero-inflation rate $\pi_0$. Because the AKSA average gives equal weight to every prefix, the influence of the spike effectively decays from full strength at $k = 1$ to zero at $k = N$, leading to an overall signal that is weaker than either the spike or the tail component taken separately.

The proposed two-step test tackles dilution by separating and then recombining the two sources of evidence. The spike test focuses exclusively on the zero spike, in which every piece of spike contrast is captured in a single mean-shift statistic, uncontaminated by observations with $X >$

0. The tail test runs on the strictly positive subset and isolates the tail signal with an AKSA-style measure that is unaffected by the mass at zero. Fisher's combination then aggregates the two one-dimensional log-likelihood contributions, effectively treating them as independent mini likelihood-ratio tests and summing their strengths. When the component tests display positive correlation, Brown's modification rescales the sum and adjusts the reference distribution, ensuring validity while retaining nearly all of the combined signal. By directing one test at the spike and another at the tail before fusing their evidence, the two-step test captures a clearer, stronger treatment-by-biomarker interaction than AKSA's single average.

## 5 Simulation Study

To validate the theoretical results, we conduct an extensive simulation study that mirrors the practical settings in predictive biomarker explorations. First, we verify that both the Fisher's and Brown's combination maintain the nominal type I error across a wide range of sample sizes and zero-inflation rates when the global null holds. Second, we show the power gain of the proposed tests over AKSA under the three alternative patterns: spike-only, tail-only, and mix. Next, we examine scenarios in which the spike and tail tests are positively correlated. We document the size inflation that arises for Fisher, show that Brown remains correctly calibrated, and compare the power of two combinations. Finally, we conduct a sensitivity analysis by introducing skewed biomarker distributions and reassess both type I error and power.

### 5.1 Control of type I error under the global null

We generate data under the global null of no treatment-by-biomarker interaction and no main-effect difference between treatment arms. Within each trial replicate, the total sample size $N \in \{60, 90, 120\}$ and the zero-inflation rate $\pi_0 \in \{0, 0.1, \ldots, 0.8\}$ are set up in a full factorial grid. A proportion $\pi_0 N$ of subjects receive $X = 0$ and the remaining $N(1 - \pi_0)$ draw $X$ from a uniform distribution $U(0,1)$. We randomly assign treatment with equal allocation, and outcomes are sampled from a standard normal distribution. For each $(N, \pi_0)$ combination, we generate 1,000 independent datasets, analyzing every replicate with AKSA, spike test, tail test, and Fisher's and Brown's combinations. Exact permutation reference distributions are obtained from 1,000 random label permutations per test, and the empirical type I error is estimated as the proportion of p-values falling below the nominal two-sided level $\alpha = 0.05$.

**Figure 1B** shows that Fisher's and Brown's combination maintains empirical type I error rates that cluster tightly around the nominal 0.05 line for every sample size and zero-inflation level examined. For $N = 60$, the point estimates oscillate by no more than 0.02 around the target. As the sample size increases to 90 and 120, the sampling variability further narrows and the two curves virtually overlap, confirming that all methods converge to the exact size in larger trials.

Importantly, we observe no systematic type I error drift with the zero-inflation rate, indicating that neither extreme sparsity nor moderate prevalence of zero biomarker values compromises type I error control when permutation calibration is used.

### 5.2 Power analysis across zero-inflation rates

For the three alternatives in power analysis, we utilize a common factorial design that varies the sample size, zero-inflation rate, and magnitude of the predictive biomarker effect. Specifically, we set the total sample size $N \in \{60, 90, 120\}$ and the zero-inflation rate $\pi_0 \in \{0, 0.1, \ldots, 0.8\}$. Each simulated dataset contains a proportion $\pi_0 N$ of subjects with $X = 0$, while the remaining $N(1 - \pi_0)$ draw their biomarker values independently from a uniform distribution $U(0,1)$. Treatment is assigned at random with equal allocation, and baseline outcomes are generated from a standard normal distribution. We simulate the biomarker predictive effect separately for the three alternatives:

**(i) Spike-only alternative.** We add a constant effect to the outcomes of treated subjects whose biomarker value is equal to zero. Mathematically, we replace the baseline outcome $Y_i$ by

$$Y_i + \Delta \times \mathbf{1}\{T_i = 1, X_i = 0\}, \quad \Delta \in \{0.8, 1, 1.2\}$$

where $T_i = 1$ indicates the subject $i$ receiving treatment and $\mathbf{1}$ is the identity function. All subjects with positive biomarker values remain unchanged.

**(ii) Tail-only alternative.** We increment the outcomes of treated subjects with positive biomarker values by a linear function of its rank. Mathematically, we replace the baseline outcome $Y_i$ by

$$Y_i + \Delta \times \mathbf{1}\{T_i = 1, X_i > 0\} \times \frac{\text{rank}(X_i)}{N}, \quad \Delta \in \{3, 4, 5\}$$

where $\text{rank}(X_i)$ is the ascending rank of the biomarker among the $N$ subjects with ties broken at random. Treated subjects at the spike receive no effect.

**(iii) Mix alternative.** We add a constant effect to treated subjects at the spike and, simultaneously, a linearly increasing effect to treated subjects with positive biomarker values. Mathematically, we replace the baseline outcome $Y_i$ by

$$Y_i + \Delta_A \times \mathbf{1}\{T_i = 1, X_i = 0\} + \Delta_B \times \mathbf{1}\{T_i = 1, X_i > 0\} \times \frac{\text{rank}(X_i)}{N},$$

where $\Delta_A \in \{0.6, 0.8\}, \Delta_B \in \{2, 3\}$.

In all three alternatives, we generate 1,000 independent trial replicates for every parameter combination. Each replicate is analyzed with AKSA, spike test, tail test, and Fisher's and Brown's combinations. Exact permutation reference distributions are obtained from 1,000 random label permutations per test. We estimate the power as the proportion of p-values that fall below the nominal two-sided level $\alpha = 0.05$.

**Figure 2** and **Supplementary Figure S1** summarize the empirical power of the competing methods when the biomarker predictive effect is confined to the zero spike. Across all combinations of effect and sample size, the two-step test dominates AKSA once the biomarker exhibits any appreciable zero inflation. For the smallest trial size $N = 60$, Fisher's combination improves power by 5-10% when $\pi_0 \leq 0.4$ and by up to 15% when $\pi_0 \geq 0.6$. Brown delivers gains of comparable magnitude while remaining slightly more conservative for the weakest effect. As total sample size increases, the absolute power of every method rises, yet the relative advantage of the two-step test remains. Even at $N = 120$, Fisher exceeds AKSA by 3-6% throughout the mid-range $\pi_0 \in [0.2, 0.6]$ and attains the 80% target earlier along the $\pi_0$ axis. Notably, both combination methods reach near-maximal power at lower zero-inflation rates than AKSA, illustrating their ability to capture spike information that the averaging step in AKSA inevitably attenuates.

**Figure 3** and **Supplementary Figure S2** show that the two-step test again outperforms under the tail-only alternative. Because no effect exists at $X = 0$, AKSA's power declines sharply as the zero-inflation rate grows, falling below 20% once $\pi_0$ exceeds 0.5 even with $N = 120$. In contrast, both Fisher and Brown remain highly sensitive. For the modest effect $\Delta = 3$, they reach or exceed the 80% benchmark at $\pi_0 \leq 0.4$ for $N = 90$ and across the entire $\pi_0$ range for $N = 120$. For larger effects $\Delta = 4$ or 5, they sustain power above 90% until the very large zero-inflation levels. Fisher and Brown curves visually converge and clearly separate from the AKSA trajectory, underscoring the benefit of isolating tail information rather than averaging it with the expanding block of zero subjects.

**Figure 4** and **Supplementary Figure S3** present that under the mixed alternative, the two-step test outperforms AKSA once $\pi_0 \geq 0.2$ across all sample sizes. Under $N = 60$ and the weaker predictive effects ($\Delta_A = 0.6$, $\Delta_B = 2$), Fisher and Brown raise power from roughly 40% to 60% at $\pi_0 = 0.5$ and the advantage widens for $N = 90$ and $N = 120$. With the stronger spike shift ($\Delta_A = 0.8$) or tail increment ($\Delta_B = 3$), the two-step test achieves or surpasses the 80% benchmark across the majority of $\pi_0$ values, whereas AKSA's power declines after peaking at moderate zero inflation. When $\pi_0$ is near zero, most subjects are in the positive tail. AKSA therefore exploits the nearly full sample with a lighter $\chi_2^2$ reference. On the other hand, the two-step test splits the data, in which the spike test is close to dormant and the tail test uses similar information but on slightly fewer subjects and then combines two components against a $\chi_4^2$ benchmark. This sample-size plus degree of freedom penalty lets AKSA retain a small edge when $\pi_0$ is close to zero, an advantage that vanishes once a meaningful spike stratum appears. Consistent with other alternatives, Fisher and Brown trajectories are virtually indistinguishable, confirming that Brown's moment-matching adjustment preserves the power gains while guarding against potential test correlation. These patterns illustrate how isolating and recombining spike and tail evidence enables the two-step test to harness both sources of signal and maintain robust sensitivity.

### 5.3 Type I error and power under correlated component tests

To examine how positive dependence between the spike and tail test affects type I error control in the two-step test, we conduct an experiment in which the two component p-values are forced to share a prespecified correlation $\rho$. For each target value $\rho \in \{0, 0.1, \dots, 0.8\}$, we generate 10,000 independent draws from a bivariate normal distribution with mean vector $(0,0)^T$ and covariance matrix $\begin{pmatrix} 1 & \rho \\ \rho & 1 \end{pmatrix}$. Transforming each component with the standard normal CDF yields a p-value pair $(p_A, p_B)$ that is marginally uniform $U(0,1)$ while exhibiting the desired correlation. The combined p-values of every replicate are calculated by Fisher's and Brown's combination, respectively. We estimate the type I error as the proportion of replicates in which the combined p-value exceeds the nominal two-sided significance level $\alpha = 0.05$.

**Figure 5A** compares the empirical type I error of Fisher's and Brown's combination under the positively correlated component tests. As correlation rises, Fisher's curve drifts upward in a roughly linear fashion, exceeding 0.07 once the correlation surpasses 0.4 and reaching almost 0.09 at $\rho = 0.8$. In contrast, Brown's moment-matching adjustment remains centered on the 0.05 benchmark across the entire $\rho$ range, with sampling error never passing beyond 0.01. These results confirm the theoretical expectation that Fisher's test becomes liberal when its independence assumption is violated, whereas Brown's scaling restores the desired type I error even under strong positive dependence.

To investigate how positive test dependence affects power, we repeat the factorial simulation in **Section 5.2** while imposing a shared random perturbation that induces a controlled correlation between the two component p-values. For each of the three alternative patterns, we set the total sample size $N \in \{60, 90, 120\}$ with the zero-inflation rate $\pi_0 \in \{0.1, 0.3, 0.5, 0.7\}$ and fix the predictive effect sizes at $\Delta_A = 0.8$ for spike-only, $\Delta_B = 2$ for tail-only, and $\Delta_A = 0.8, \Delta_B = 2$ for mix. Within every trial replicate, a non-negative normal draw $Z \sim |N(0,1)|$ is added, then scaled by a factor $k \in \{0.5, 1, 2, 3\}$ to the outcomes of all treated subjects, thereby generating approximate correlations between $p_A$ and $p_B$ ranging from 0.1 to 0.5. We simulate 1,000 independent trials for every $(N, \pi_0, k)$ combination. Brown's combination is applied at its nominal two-sided level $\alpha = 0.05$ while Fisher's critical value is selected by the correlation-specific threshold that yields a 0.05 type I error under the previous dependent-null experiment. We calculate the empirical power as the proportion of replicates in which the resulting p-value falls below respective critical threshold.

**Figure 5B** shows the power difference of Fisher's and Brown's combination averaging across simulation replicates. The power difference remains virtually indistinguishable among most correlation between the component p-values. The difference never exceeds 0.02 and oscillates evenly around zero, confirming that Brown's scaling does not erode sensitivity in practical settings. A modest positive edge for Brown appears at $\rho \approx 0.2$ under the smallest trial size $N = 60$, but the advantage vanishes as $N$ grows or as $\rho$ increases further. These patterns demonstrate that Brown remains power under the positive test correlation without sacrificing the protection against type I error.

**5.4 Robustness to right-skewed biomarker distributions**

Right-skewed biomarker distributions arise frequently in clinical and translational studies [27,28]. To assess whether it compromises type I error control, we repeat the null experiment of **Section 5.1** while replacing the uniform positive tail with three increasingly skewed Beta distributions. For each trial replicate, we first fix the total sample size at $N \in \{60, 90, 120\}$ and select a zero-inflation rate $\pi_0 \in \{0, 0.1, \ldots, 0.8\}$. A proportion $\pi_0 N$ of subjects are assigned $X = 0$ and the remaining $N(1 - \pi_0)$ receive a strictly positive biomarker drawn independently from Beta(2, 5) (mild skewness), Beta(1, 4) (moderate skewness), or Beta(0.5, 3) (strong skewness). Treatment is randomized at 1:1 and outcomes are sampled from a standard normal distribution, ensuring the global null of no treatment-biomarker interaction and no main effect. We analyze each dataset by all competing methods, using 1,000 random label permutations per test to obtain exact reference distributions. The simulation comprises 1,000 independent replicates for every ($N$, $\pi_0$, skewness) combination, and empirical type I error is estimated as the proportion of p-values that fall below the nominal two-sided level $\alpha = 0.05$.

**Supplementary Figures S4-S6** depict the empirical type I error of AKSA, Fisher's, and Brown's combination under a right-skewed biomarker. Across all three skewness settings, the point estimates cluster tightly around the nominal 0.05 line for every sample size and zero-inflation rate examined. Importantly, neither the heavier tail nor the excess of extreme biomarker values induces any systematic inflation. Both Fisher and Brown stay indistinguishable from AKSA, oscillating by at most 0.02 with no discernible upward or downward drift as $\pi_0$ grows from 0 to 0.8. These results confirm that the proposed two-step test continues to deliver exact type I error control even when the biomarker distribution departs markedly from uniformity.

To study whether the two-step test preserves its power advantage under skewness, we rerun the three alternative simulations of **Section 5.2** with the positive biomarker tail drawn from right-skewed Beta distributions. For each alternative, we consider three degrees of skewness defined in the previous null study. Within every replicate, the total sample size is fixed at $N \in \{60, 90, 120\}$, the zero-inflation rate at $\pi_0 \in \{0, 0.1, \ldots, 0.8\}$, and the predictive effect sizes at $\Delta = 0.8$ for spike-only, $\Delta = 3$ for tail-only, and $\Delta_A = 0.6$, $\Delta_B = 2$ for mix. After randomly assigning treatment with equal allocation, outcomes are generated from a standard normal distribution and then perturbed according to the specified alternative. We analyze each dataset by all competing methods. The p-value distributions are obtained from 1,000 random label permutations, and power is estimated as the proportion of p-values falling below the $\alpha = 0.05$ from 1,000 independent trial replicates.

**Supplementary Figures S7-S12** compare the empirical power of the two-step test with AKSA when the positive tail of the biomarker is right-skewed. Under the spike-only alternative, Fisher and Brown consistently outperform AKSA across all zero-inflation rates and sample sizes, gaining 3%-7% even when the skew was strong. Tail-only scenarios show a larger separation: once $\pi_0$ exceeds 0.2, the averaging step in AKSA dilutes the monotone signal, whereas the two-step test preserves ≥ 75% power for $N \geq 90$ and maintains an 80% benchmark up to $\pi_0 \approx 0.6$. For the mix alternative, the two-step curves again rise above those of AKSA throughout, achieving the 80% target earlier and retaining 10%-15% absolute advantage at moderate skewness. Importantly, the three skewness settings yield virtually parallel trajectories, indicating the power robustness of Fisher's log-likelihood summation and Brown's moment-matching adjustment. The principal determinant of power remains the degree to which the spike component is isolated rather than averaged away.

# 6 Discussion

Our study carries important practical implications for early-phase clinical trials in precision medicine. Early-phase trials are typically small and exploratory, yet decisions made at this stage can shape later-phase development [29]. The proposed two-step test provides a rigorous but flexible tool for detecting predictive biomarkers with limited sample sizes and complex biomarker distributions. In practice, biomarkers such as PD-L1 expression in immunotherapy trials often exhibit a point mass of zero values alongside a continuous range of positive values [30,31]. In this scenario, our test explicitly handles the zero-inflation and skewness, making early-phase trials reliably assess a biomarker's predictive value without being misled by the large spike of zero readings. With power to detect true predictive signals while controlling false positives, our method can help researchers decide which biomarkers merit validation in larger trials, ultimately accelerating the development of targeted therapies.

Our two-step test shows improved power compared to the recent AKSA method across a range of scenarios, alongside reliable type I error control. AKSA is inspired by the KS distance that was proposed to assess biomarker predictiveness in small-sample trials. Prior work has shown that AKSA outperforms classical tests when the biomarker's predictive effect follows a complex nonlinear pattern [10]. However, AKSA does not explicitly account for a point mass at zero, which may dilute signals that occur in a subrange of that distribution. Our two-step test excels in situations where the predictive effect manifests in either the spike or the tail in isolation. More importantly, these power gains come with exact type I error control by creating a unified permutation null distribution and appropriate combination methods. The two-step test learns where the treatment-biomarker interaction lies while still correctly declaring no significance when there is no effect.

A key strength of our approach is its nonparametric and permutation-calibrated nature, which makes the test inherently robust to zero-inflation and skewness. Our two-step test makes minimal assumptions without assuming a specific outcome model or biomarker distribution. Each component test is evaluated via permutation using the observed data to simulate the null distribution of the test statistic. This guarantees valid type I error control at the nominal level without relying on large-sample theory or asymptotic distributions [32]. Such permutation calibration is particularly valuable in small-sample trials, where standard asymptotic p-values can be inaccurate [33]. An additional advantage of our method is flexibility. It accommodates different test statistics and can easily incorporate alternative combination rules if needed [34]. The procedure naturally reduces to a standard single-part test when the biomarker has no zero inflation or no continuous tail [35]. In those situations, our two-step method focuses on whichever component is applicable, incurring no loss of power relative to an optimal one-part test.

The proposed two-step test has several limitations and practical considerations. First, the approach partitions the data by biomarker values, which can be viewed as a form of data splitting. If the proportion of biomarker-positive subjects is low, the continuous part of the analysis may be under power, and it is similar for the spike test. Researchers should be aware of this, especially in small trials or highly imbalanced biomarker distributions. We recommend reporting the results of both subtests alongside the combined result to show if one part is driving the significance or has very low sample size. This transparency will help interpretation and understanding the nature of the predictive effect. Another consideration is the estimation of correlation between component tests. In practice, one can estimate the correlation empirically by the permutation replicates and plug that into Brown's combination formula. However, in very small samples, or if the test statistics have a non-normal joint distribution, the estimation might not be accurate. Alternatively, one could generate a null distribution of the combined statistic by permuting treatment labels and computing the two-step statistic each time [36]. This approach would inherently account for any dependence and yield an exact p-value under the permutation null. The tradeoff is computational cost, especially if a large number of permutations are required to get a stable p-value for the combined test [37,38].

There are several directions to extend the current method. First, many trials collect subject covariates, and it would be useful to adjust the biomarker test for these variables. Future work could integrate covariate adjustment by using stratified permutation schemes [39,40]. For example, one could permute treatment labels within strata defined by important prognostic variables to account for their effect, ensuring that the biomarker's predictiveness is assessed without confounders. Second, a next step is adapting the two-part test to time-to-event outcomes. Permutation approaches can be applied in this context by permuting survival outcomes. Our framework can accommodate such endpoints using a log-rank test for the tail component and combining it with a test for survival difference between biomarker-negative and positive patients [41,42]. Third, researchers often in practice examine panels of candidate biomarkers or high-dimensional omics data to find predictive signals. One direction would be generalizing the two-step test to multiple features, separating tests for a multivariate spike and for multivariate differences in the distribution of a composite score. Alternatively, one could apply our two-step test to each biomarker in a panel and adjust for multiple comparisons to identify a set of predictive biomarkers [43,44]. Bringing the power of our test to multi-biomarker signatures could enhance early-phase precision medicine studies, where often no single biomarker is sufficient but a combination is predictive.

In conclusion, our two-step test represents a statistically innovative and practically relevant advancement for early-phase trial analysis. By addressing the challenges of zero-inflated biomarker distributions, it enables more reliable detection of predictive biomarkers, ultimately aiding the design of personalized therapies. This approach serves as a foundation for robust biomarker inference, and its application will benefit precision medicine.

## Data Availability Statement

All materials needed to reproduce the results of this study are openly accessible below:

**R package** "twostepAKSA" – source code and documentation are available from

https://github.com/xnnba1984/twostepAKSA.

**Analysis code** – contained in the companion repository

https://github.com/xnnba1984/Two_step_test_paper.

## Conflicts of Interest

Nan Miles Xi are full time employees of AbbVie. AbbVie had no role in the design, analysis, interpretation, or decision to publish this study. All other authors declare no conflicts of interest.

## Funding

Lin Wang is supported by the National Science Foundation (DMS-2413741) and the Central Indiana Corporate Partnership AnalytiXIN Initiative.

## Author Contributions

# Figures

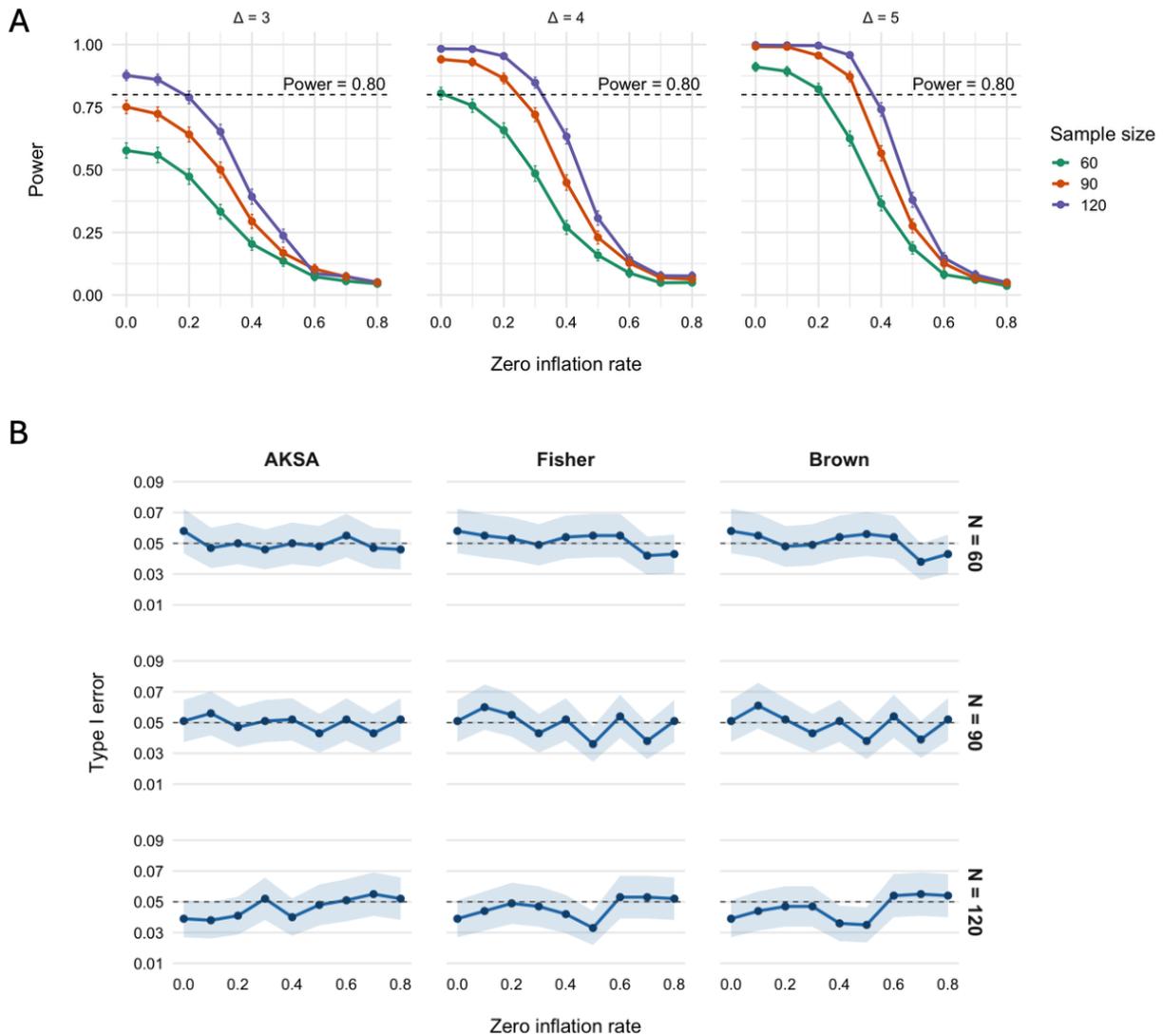

**Figure 1. AKSA power loss and type I error control of the proposed two-step test. (A) Power loss of AKSA under the tail-only alternative.** Curves represent the empirical power of AKSA for sample sizes 60 (green), 90 (orange) and 120 (purple) as the zero-inflation rate increases from 0 to 0.8, with columns corresponding to monotone effect slopes Δ = 3, 4, 5. The dashed horizontal line marks the 80% target. **(B) Type I error control at the global null.** Empirical rejection rates for AKSA, Fisher's and Brown's combinations are shown in separate columns with rows corresponding to sample sizes 60, 90, 120. Shaded ribbons denote Monte-Carlo 95% confidence intervals based on 1,000 simulated trials per design point, and the dashed line indicates the nominal level $\alpha = 0.05$.

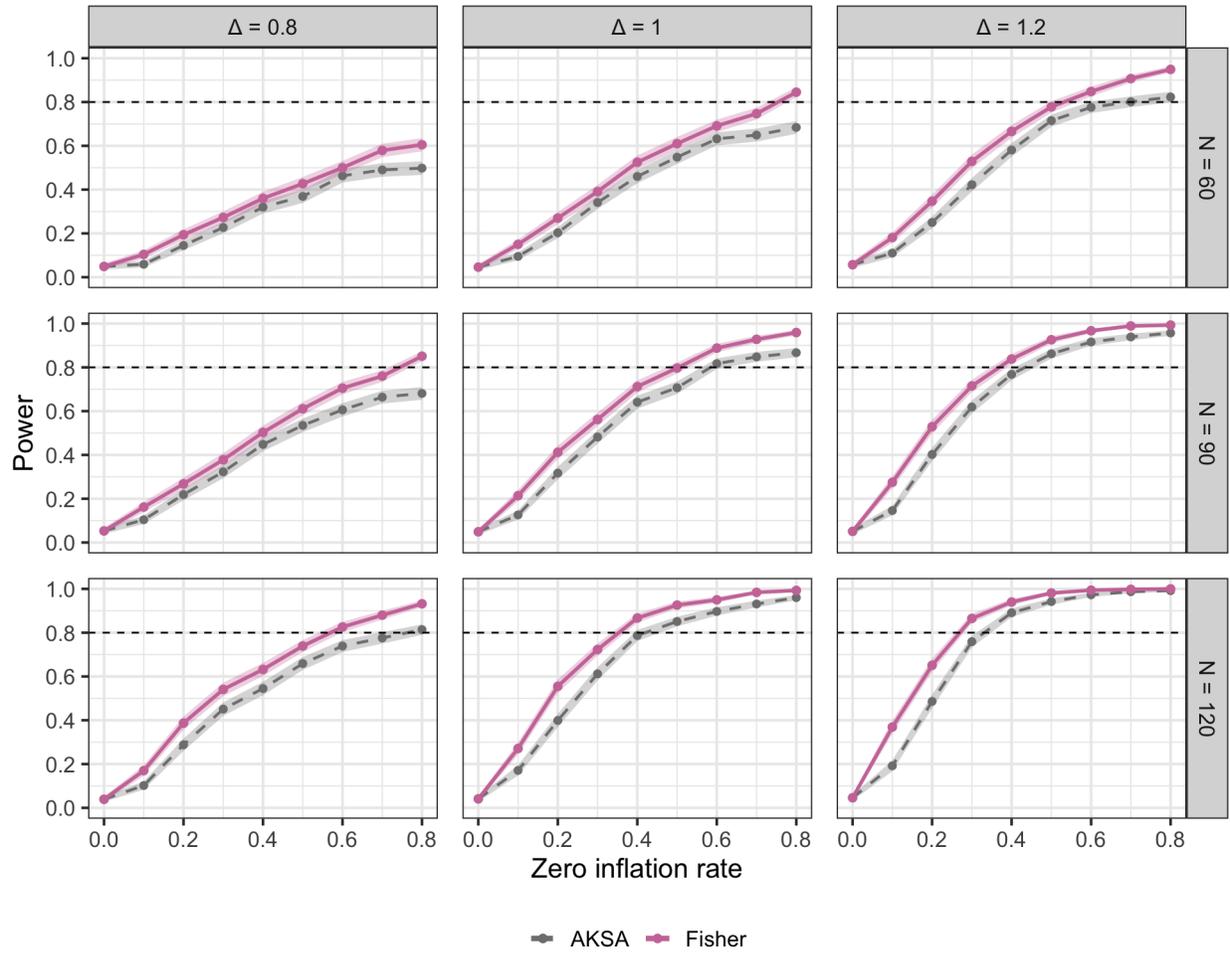

**Figure 2. Empirical power of AKSA and Fisher's combination under the spike-only alternative.** Each panel displays the proportion of 1,000 simulated trials in which the corresponding test rejected the global null (y-axis) as the zero-inflation rate varied from 0 to 0.8 (x-axis). Columns distinguish the size of the treatment effect at the spike and rows represent sample size. Grey curves depict AKSA and magenta curves depict the two-step test combined by Fisher's method. Shaded ribbons are Monte-Carlo 95% confidence intervals, and the horizontal dashed line marks the 80% power benchmark.

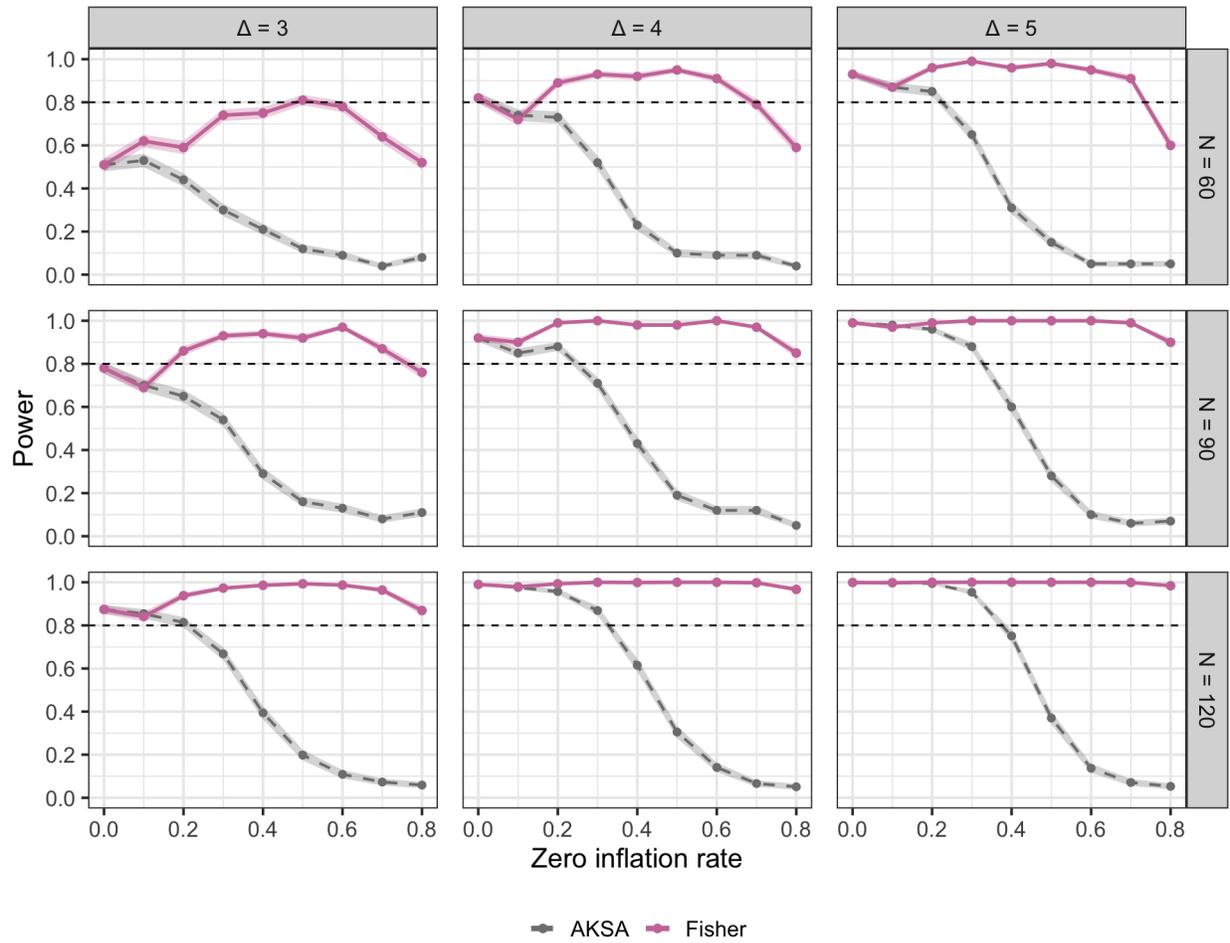

**Figure 3. Empirical power of AKSA and Fisher's combination under the tail-only alternative.** Each panel displays the proportion of 1,000 simulated trials in which the corresponding test rejected the global null (y-axis) as the zero-inflation rate varied from 0 to 0.8 (x-axis). Columns distinguish the size of the treatment effect at the tail, while rows represent sample size. Grey curves depict AKSA and magenta curves depict the two-step test combined by Fisher's method. Shaded ribbons are Monte-Carlo 95 % confidence intervals, and the horizontal dashed line marks the 80% power benchmark.

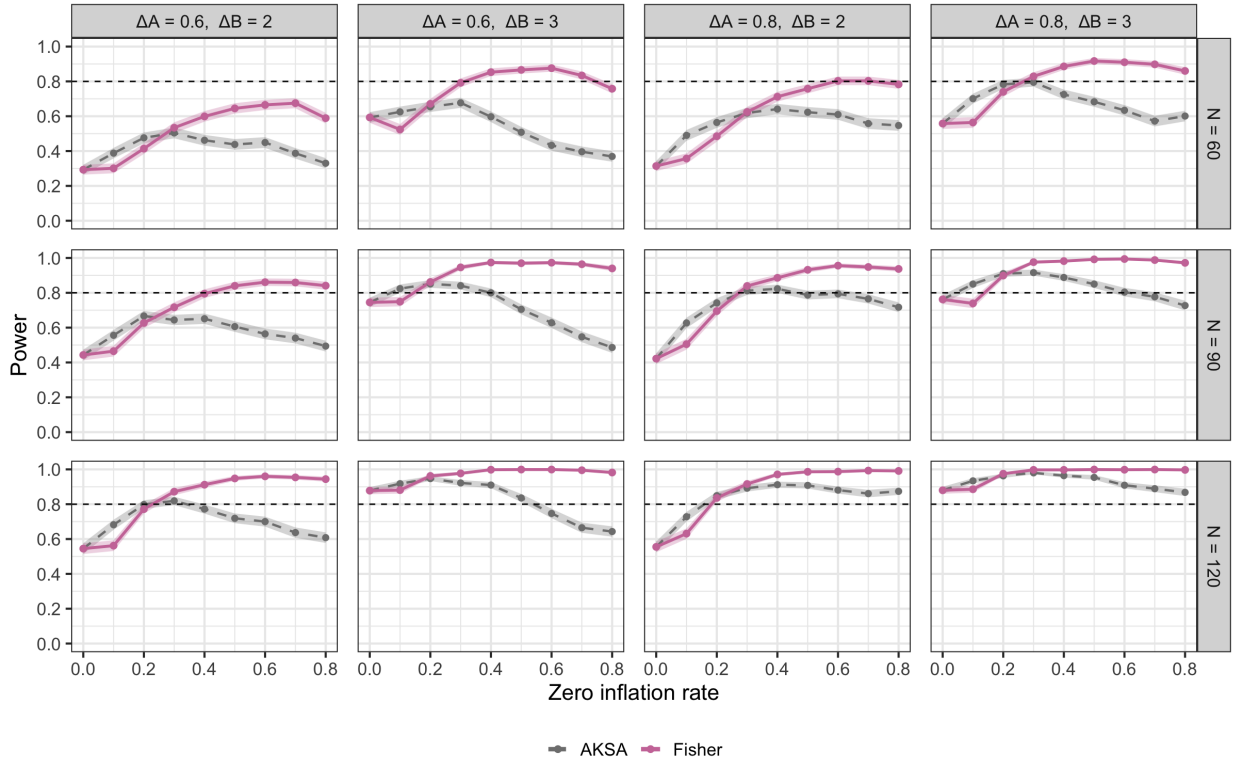

**Figure 4. Empirical power of AKSA and Fisher's combination under the mix alternative.** Each panel displays the proportion of 1,000 simulated trials in which the corresponding test rejected the global null (y-axis) as the zero-inflation rate varied from 0 to 0.8 (x-axis). Columns distinguish the size of the treatment effect, while rows represent sample sizes. Grey curves depict AKSA and magenta curves depict the two-step test combined by Fisher's method. Shaded ribbons are Monte-Carlo 95 % confidence intervals, and the horizontal dashed line marks the 80% power benchmark.

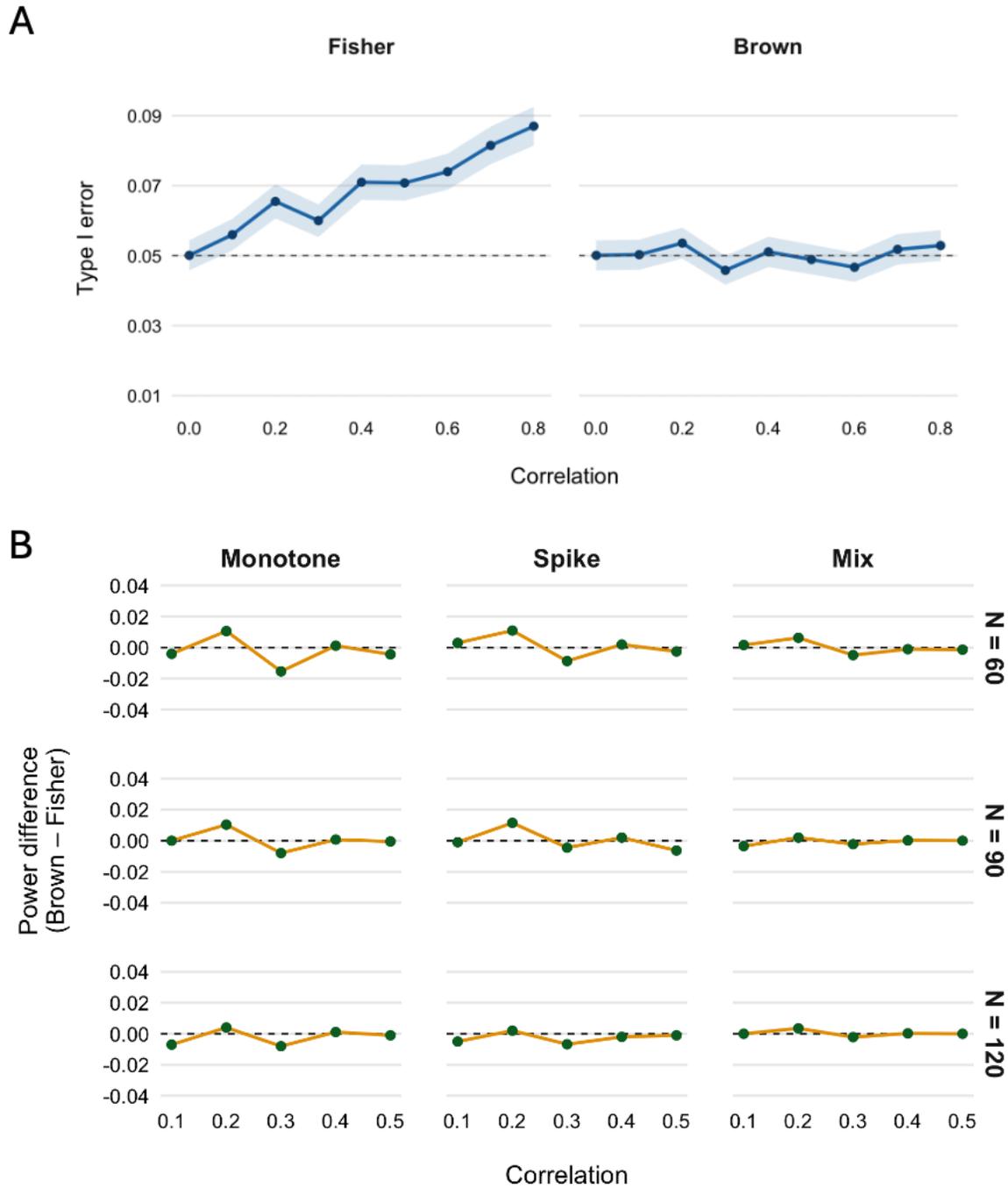

**Figure 5. Impact of dependence between spike and tail test on type I error and power. (A)** Empirical type I error of Fisher's (left) and Brown's combination (right) when the two component p-values are sampled from a bivariate normal copula with marginal $U(0,1)$ distributions and target correlations under the global null. Points denote Monte-Carlo estimates from 10,000 replicates and shaded bands give 95 % confidence intervals. **(B)** Power difference under alternatives between Fisher and Brown. Each point corresponds to 1,000 simulated trials across zero-inflation rates.

# Appendix

## A.1 Power Loss of AKSA in the Presence of Zero-Inflated Biomarkers

### A.1.1 AKSA statistic

Denote the true log-odds treatment effect curve at biomarker value $x$ as

$$D(x) = f_1(x) - f_0(x)$$

where

$$f_k(x) = \text{logit}\Pr\{Y = 1 \mid T = k, X = x\}, \ k \in \{0,1\}$$

AKSA estimates $D(x)$ from a fitted regression model, draws $B$ pairs $(x_1, x_2)$ from the empirical biomarker distribution, and forms

$$d_i = |\widehat{D}(x_2) - \widehat{D}(x_1)| + \varepsilon_i, \qquad i = 1, \ldots, B$$

with $\varepsilon_i \sim N(0, \sigma_i^2)$. The test statistic is the Monte-Carlo estimate

$$\widehat{P} = \frac{1}{B} \sum_{i=1}^{B} \mathbf{1}\{d_i > 0\}$$

We reject $H_0: E_X[D(X)] = 0$ whenever $\widehat{P} > \alpha_{\text{AKSA}}$. Note that $\widehat{P}$ is an increasing transform of an asymptotically normal $Z$ statistic whose noncentrality parameter is proportional to $\Delta$ [1]. Therefore, for large $N$ the sampling distribution of $\widehat{P}$ is driven by the population average treatment effect

$$\Delta = E_X[D(X)]$$

and the power of AKSA is essentially a function of $\Delta$.

### A.1.2 Average treatment effect under zero inflation

Assume the biomarker has a point mass at zero:

$$X \sim \pi_0 \delta_0 + (1 - \pi_0) H \tag{1}$$

where $0 \leq \pi_0 < 1$, $\delta_0$ is the Dirac measure at 0, and $H$ is a continuous distribution on $(0, +\infty)$. Write expectations under $H$ as $E_H[\cdot]$. Partitioning on $X = 0$ versus $X > 0$ in (1) gives

$$\Delta(\pi_0) = \pi_0 D(0) + (1 - \pi_0) E_H[D(X)] \tag{2}$$

Define the baseline average treatment effect (no zero inflation) as

$$\Delta_0 := \Delta(0) = E_H[D(X)] \qquad (3)$$

We make the monotone-effect assumption: $D(x)$ is non-decreasing in $x$ and strictly greater for some $x > 0$. In practice, this corresponds to a biomarker whose higher values enhance the treatment effect. Note that AKSA itself is shape-agnostic and can be paired with linear or nonparametric fits. For analytic tractability, we state our power result under a monotone-effect condition, which is commonly met in practice but not required by the original AKSA.

Because $D(x)$ is increasing, we have

$$D(0) < E_H[D(X)] = \Delta_0$$

Plugging in (3) into (2) gives

$$\Delta(\pi_0) = \Delta_0 - \pi_0(\Delta_0 - D(0))$$

Hence $\Delta(\pi_0)$ is strictly decreasing in $\pi_0$ and

$$\Delta(\pi_0) < \Delta_0 \text{ for every } \pi_0 > 0$$

### A.1.3 Power loss result

Let $\widehat{\Delta}$ be a $\sqrt{N}$-consistent estimator of $\Delta$ under alternative with $\text{Var}(\widehat{\Delta}) = \sigma^2/N + o(N^{-1})$. The core $Z$-statistic driving $\widehat{P}$ is

$$Z_{\text{AKSA}} = \frac{\sqrt{N}\widehat{\Delta}}{\sigma} = \frac{\sqrt{N}(\widehat{\Delta} - \Delta(\pi_0))}{\sigma} + \frac{\sqrt{N}\Delta(\pi_0)}{\sigma}$$

Under large sample, the first term turns to $N(0,1)$ by the Central Limit Theorem. The second term is constant for a fixed $\pi_0$. Therefore,

$$Z_{\text{AKSA}} \xrightarrow{d} N(\lambda(\pi_0), 1), \quad \lambda(\pi_0) = \frac{\sqrt{N}\Delta(\pi_0)}{\sigma} \qquad (4)$$

Since $\Delta(\pi_0)$ is strictly decreasing in $\pi_0$ while $\sigma$ is unaffected to first order, the noncentrality parameter $\lambda(\pi_0)$ is also strictly decreasing in $\pi_0$. For the AKSA test at level $\alpha$, the asymptotic power is

$$\beta_{\text{AKSA}}(\pi_0) = 1 - \Phi\left(z_{1-\alpha/2} - \lambda(\pi_0)\right) + \Phi\left(-z_{1-\alpha/2} - \lambda(\pi_0)\right)$$

Take derivative of $\beta_{\text{AKSA}}(\pi_0)$ with respect to $\lambda(\pi_0)$ gives

$$\frac{d\beta_{\text{AKSA}}(\pi_0)}{d\lambda(\pi_0)} = \phi(z_{1-\alpha/2} - \lambda(\pi_0)) - \phi(-z_{1-\alpha/2} - \lambda(\pi_0)) > 0$$

The inequality holds if $\lambda(\pi_0) > 0$, which is true under the alternative. Therefore, $\beta_{AKSA}(\pi_0)$ is strictly increasing in $\lambda(\pi_0)$ and

$$\pi_0^{(1)} > \pi_0^{(2)} \implies \lambda(\pi_0^{(1)}) < \lambda(\pi_0^{(2)}) \implies \beta_{AKSA}(\pi_0^{(1)}) < \beta_{AKSA}(\pi_0^{(2)})$$

That is, the power of AKSA is negatively related to zero inflation rate $\pi_0$.

## A.2 Type I Error Control of Fisher's Combination under Independent Tests

### A.2.1 Distribution of the Fisher statistic

Let $p_A$ and $p_B$ denote the p-values produced by the spike test and tail test. Under the global null hypothesis

$$H_0: E_X[D(X)] = 0$$

$p_A$ and $p_B$ are both from a uniform distribution $U(0,1)$. Further assume $p_A$ and $p_B$ are independent (i.e., the spike and tail tests are independent). For any $j \in \{A, B\}$, let $U_j = -2\log p_j$. Because $p_j \sim U(0,1)$, we have

$$\Pr\{U_j \leq u\} = \Pr\{-2\log p_j \leq u\} = \Pr\left\{p_j \geq e^{-\frac{u}{2}}\right\} = 1 - e^{-\frac{u}{2}}, \quad u \geq 0 \quad (5)$$

The rightmost expression is the CDF of $\chi_2^2$, so $U_j \sim \chi_2^2$. Since the monotone transformation $u \to -2\log u$ preserves independence, $U_A$ and $U_B$ are also independent. Note that the sum of two independent $\chi_2^2$ is $\chi_4^2$, hence we have

$$S_{Fisher} = U_A + U_B \sim \chi_4^2 \quad (6)$$

### A.2.2 Uniformity of the Fisher p-value

The p-value of Fisher's combination is computed as

$$p_{Fisher} = \Pr\{\chi_4^2 \geq S_{Fisher}\} = 1 - F_{\chi_4^2}(S_{Fisher})$$

where $F_{\chi_4^2}$ is the CDF of $\chi_4^2$. By the probability integral transform, for any continuous random variable $Z$ with CDF $F_Z$, the transformed variable $U = F_Z(Z)$ follows a uniform distribution $U(0,1)$. Equivalently, $1 - F_Z(Z)$ is also $U(0,1)$ because $1 - U$ has the same distribution as $U$. Apply this to $Z = S_{Fisher}$ gives

$$p_{Fisher} = 1 - F_{\chi_4^2}(S_{Fisher}) \sim U(0,1), \quad \Pr\{p_{Fisher} \leq \alpha\} = \alpha, \quad \forall 0 < \alpha < 1,$$

That is, the null distribution of $p_{Fisher}$ is uniform and the test that rejects when $p_{Fisher} \leq \alpha$ satisfies

$$\Pr_{H_0}\{\text{reject } H_0\} = \alpha$$

Therefore, the Fisher combination exactly controls the type I error at level $\alpha$.

## A.3 Type I Error Control of Brown's Combination under Correlated Tests

### A.3.1 Moments of the Fisher statistic

Using the same notation in **Section A.2** and under $H_0$, $p_A$ and $p_B$ are from a uniform distribution $U(0,1)$. The transformed variables

$$U_A = -2\log p_A, \qquad U_B = -2\log p_B$$

have finite covariance

$$\text{Cov}(U_A, U_B) = \sigma_{AB} \geq 0$$

For $j \in \{A, B\}$, denote $\mu_j = E(U_j)$ and $\sigma_j^2 = \text{Var}(U_j)$. Recall that Equation (5) gives

$$\Pr\{U_j \leq u\} = 1 - e^{-\frac{u}{2}}, \qquad u \geq 0$$

which is the CDF of an exponential distribution with $\lambda = \frac{1}{2}$. Therefore,

$$\mu_j = \frac{1}{\lambda} = 2, \qquad \sigma_j^2 = \frac{1}{\lambda^2} = 4$$

Write the expectation and variance of Fisher's test statistics as $\mu_S$ and $\sigma_S^2$, by Equation (6) we have

$$\mu_S = \mu_A + \mu_B = 4, \qquad \sigma_S^2 = \sigma_A^2 + \sigma_B^2 + 2\sigma_{AB} = 8 + 2\sigma_{AB} = (1 + \rho) \tag{7}$$

### A.3.2 Moments matching with the Brown statistic

The aim of Brown's combination is to replace $S_{\text{Fisher}}$ by a scaled chi-square random variable $\chi_\nu^2$ whose first two moments coincide with those in (7). To achieve this, let $S_{\text{Brown}} = \frac{S_{\text{Fisher}}}{c} \approx \chi_\nu^2$, and note that

$$E(S_{\text{Brown}}) = E(\chi_\nu^2) = \nu, \qquad \text{Var}(S_{\text{Brown}}) = \text{Var}(\chi_\nu^2) = 2\nu$$

Matching first and second moments gives

$$c\nu = \mu_S, \qquad 2c^2\nu = \sigma_S^2 \tag{8}$$

Solving Equation (8) by Equation (7) gives

$$c = 1 + \rho, \qquad v = \frac{4}{c}$$

With such values of $c$ and $v$, $S_{\text{Brown}}$ and $\chi_v^2$ have the identical first and second moments. Because $c \geq 1$ and $v \leq 4$ whenever $\rho \geq 0$, the random variable $c\chi_v^2$ is a mean-preserving spread of $S_{\text{Fisher}} = \chi_4^2$, and its distribution is heavier-tailed than that of $S_{\text{Fisher}} = \chi_4^2$ in the convex-order sense. At the same time, $S_{\text{Fisher}}$ itself is a sum of positively correlated, gamma-distributed components, and classical results on association imply $S_{\text{Fisher}}$ is stochastically no larger than $c\chi_v^2$ [2]. Since $S_{\text{Brown}} = \frac{S_{\text{Fisher}}}{c}$ and dividing by the positive constant $c$ preserves this order, we have

$$S_{\text{Brown}} \preceq_{st} \chi_v^2 \tag{9}$$

### A.3.3 Super-uniform of the Brown p-value

Let $F_v$ denote the CDF of $\chi_v^2$. Relation (9) implies

$$F_v(x) \leq \Pr\{S_{\text{Brown}} \leq x\}, \qquad \forall x \geq 0$$

Consequently,

$$p_{\text{Brown}} = \Pr\{\chi_v^2 \geq S_{\text{Brown}}\} = 1 - F_v(S_{\text{Brown}}) \geq 1 - \Pr\{S_{\text{Brown}} \leq S_{\text{Brown}}\}$$

By probability integral transform, $1 - \Pr\{S_{\text{Brown}} \leq S_{\text{Brown}}\}$ follows a uniform distribution $U(0,1)$. Therefore, $p_{\text{Brown}}$ is stochastically larger than a uniform random variable and super-uniform. For any $0 < \alpha < 1$,

$$\Pr_{H_0}\{p_{\text{Brown}} \leq \alpha\} \leq \Pr\{U \leq \alpha\} = \alpha$$

That is, the Brown's combination controls the type I error at level $\alpha$.

# Supplementary Figures

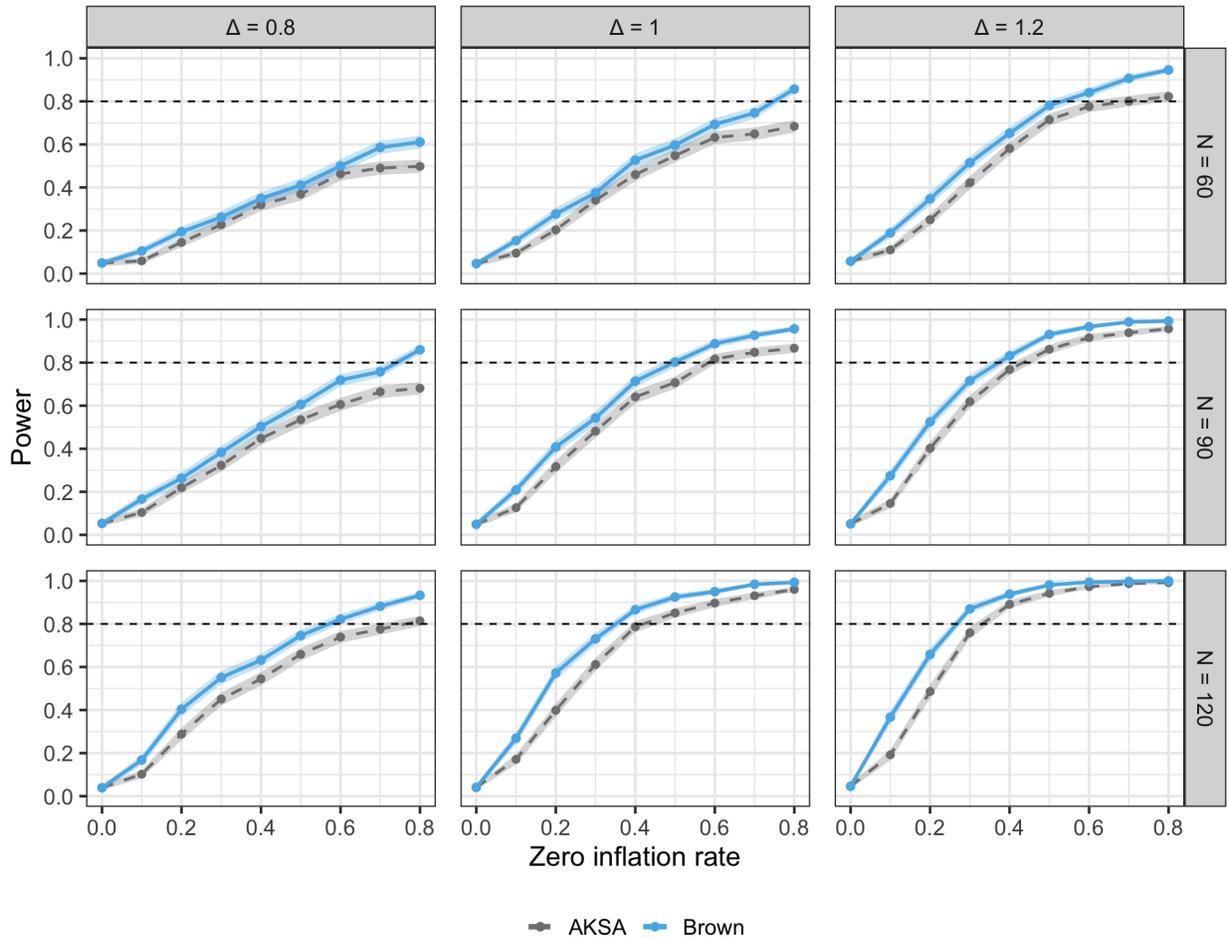

**Supplementary Figure S1. Empirical power of AKSA and Brown's combination under the spike-only alternative.** Each panel displays the proportion of 1,000 simulated trials in which the corresponding test rejected the global null (y-axis) as the zero-inflation rate varied from 0 to 0.8 (x-axis). Columns distinguish the size of the treatment effect at the spike, while rows represent sample sizes. Grey curves depict AKSA and magenta curves depict the two-step test combined by Brown's method. Shaded ribbons are Monte-Carlo 95 % confidence intervals, and the horizontal dashed line marks the 80% power benchmark.

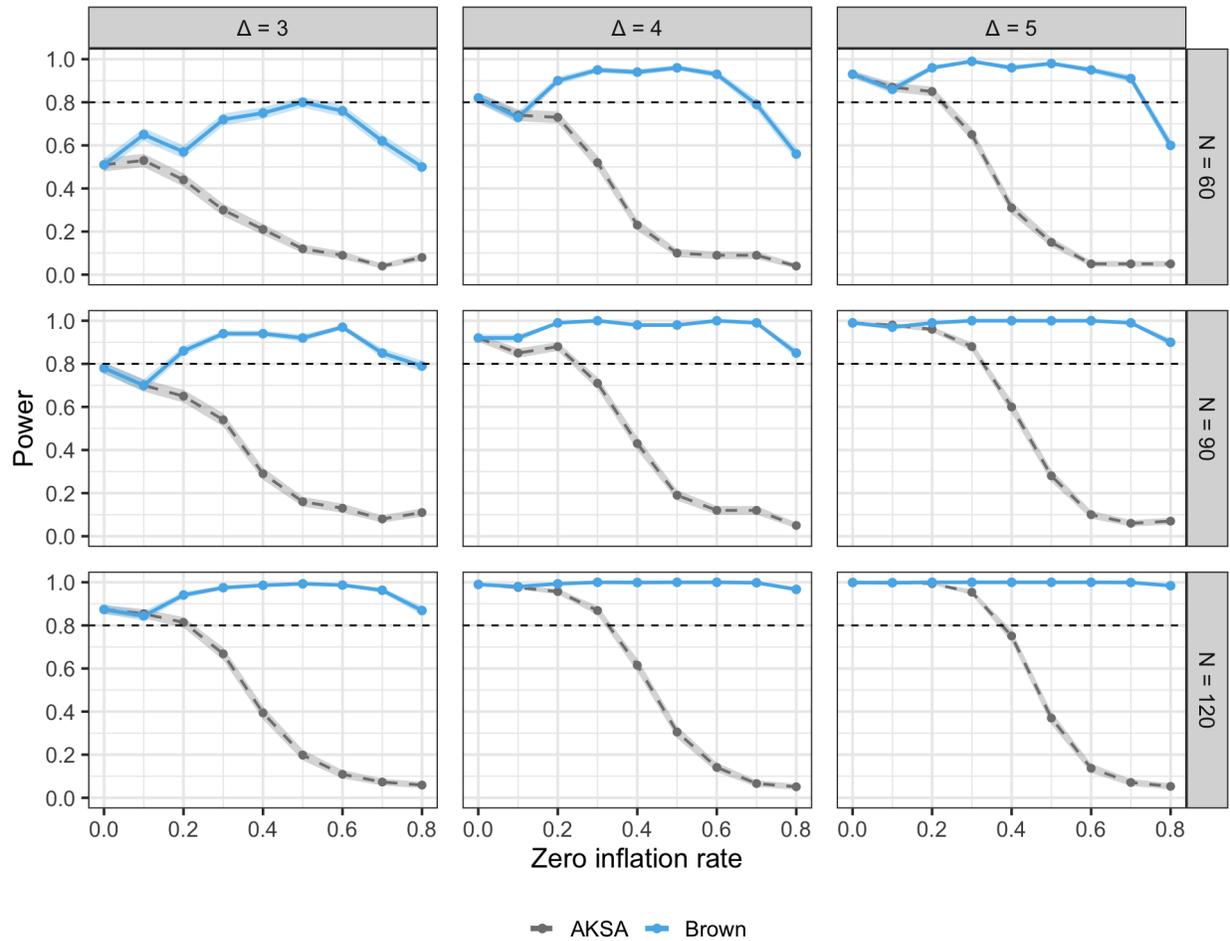

**Supplementary Figure S2. Empirical power of AKSA and Brown's combination under the tail-only alternative.** Each panel displays the proportion of 1,000 simulated trials in which the corresponding test rejected the global null (y-axis) as the zero-inflation rate varied from 0 to 0.8 (x-axis). Columns distinguish the size of the treatment effect at the spike, while rows represent sample sizes. Grey curves depict AKSA and magenta curves depict the two-step test combined by Brown's method. Shaded ribbons are Monte-Carlo 95 % confidence intervals, and the horizontal dashed line marks the 80% power benchmark.

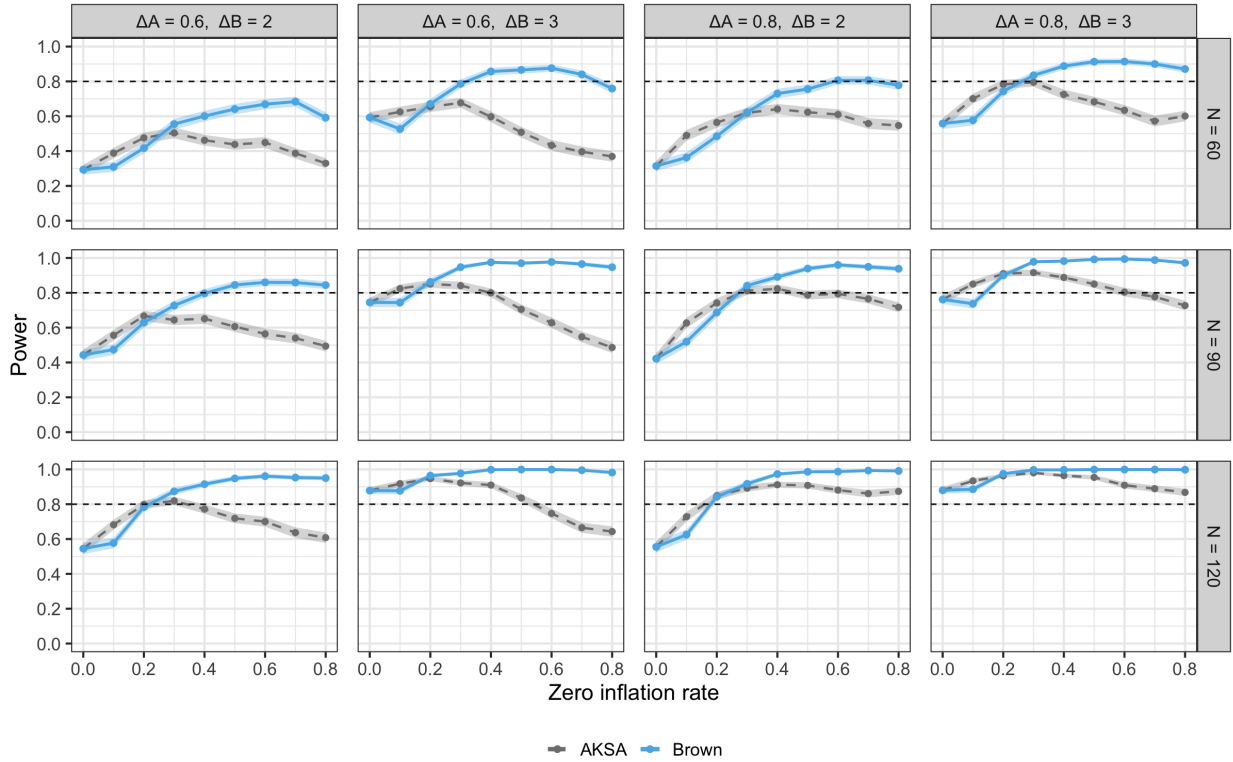

**Supplementary Figure S3. Empirical power of AKSA and Brown's combination under the mix alternative.** Each panel displays the proportion of 1,000 simulated trials in which the corresponding test rejected the global null (y-axis) as the zero-inflation rate varied from 0 to 0.8 (x-axis). Columns distinguish the size of the treatment effect, while rows represent sample sizes. Grey curves depict AKSA and magenta curves depict the two-step test combined by Brown's method. Shaded ribbons are Monte-Carlo 95 % confidence intervals, and the horizontal dashed line marks the 80% power benchmark.

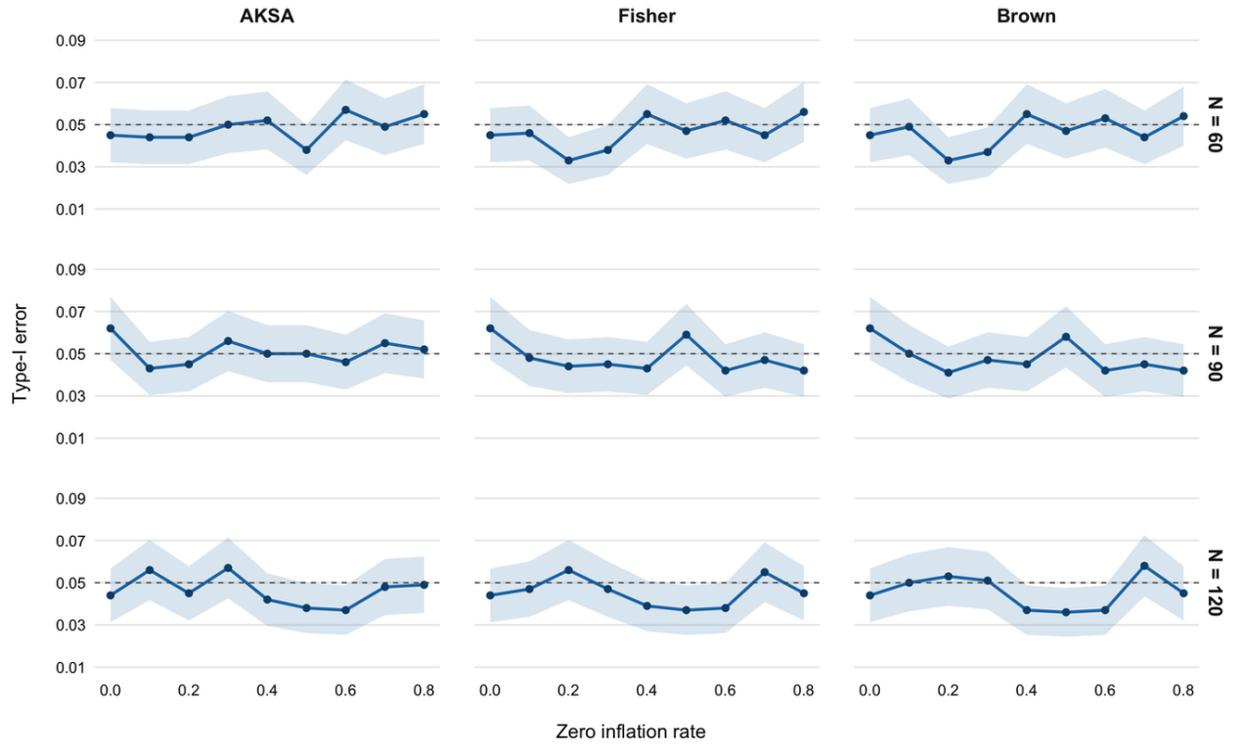

**Supplementary Figure S4. Type I error control of competing tests with a mildly skewed biomarker.** Empirical rejection rates for AKSA, Fisher's and Brown's combinations are shown in separate columns with rows corresponding to sample sizes 60, 90, 120. Shaded ribbons denote Monte-Carlo 95% confidence intervals based on 1,000 simulated trials per design point, and the dashed line indicates the nominal level $\alpha = 0.05$. Biomarker values are drawn from a Beta(2, 5) distribution indicating mild skewness.

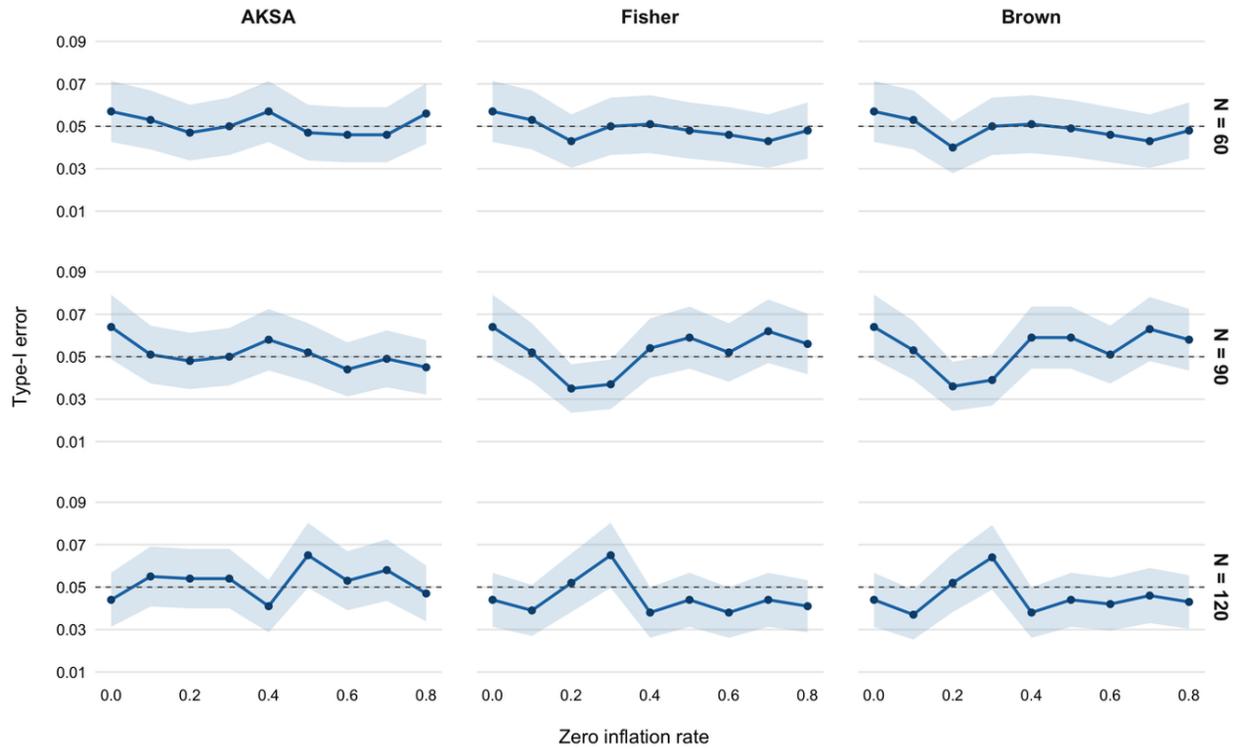

**Supplementary Figure S5. Type I error control of competing tests with a moderately skewed biomarker.** Empirical rejection rates for AKSA, Fisher's and Brown's combinations are shown in separate columns with rows corresponding to sample sizes 60, 90, 120. Shaded ribbons denote Monte-Carlo 95% confidence intervals based on 1,000 simulated trials per design point, and the dashed line indicates the nominal level $\alpha = 0.05$. Biomarker values are drawn from a Beta(1, 4) distribution indicating moderate skewness.

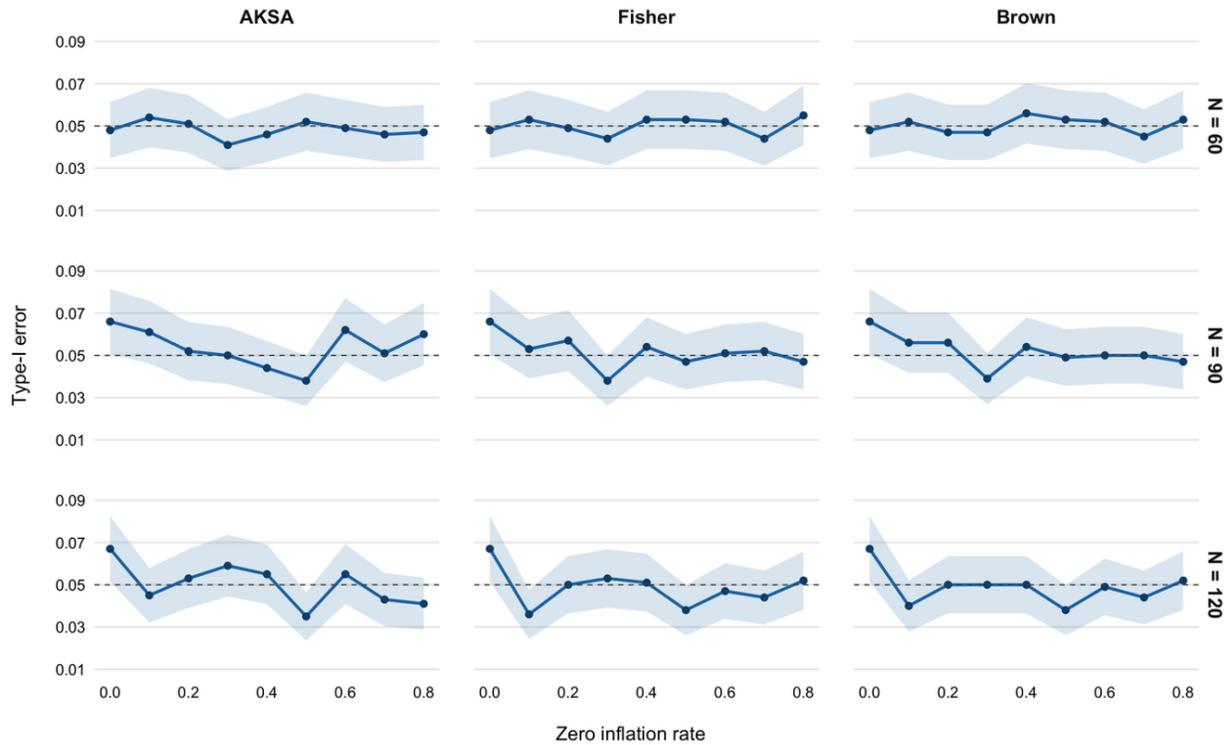

**Supplementary Figure S6. Type I error control of competing tests with a strongly skewed biomarker.** Empirical rejection rates for AKSA, Fisher's and Brown's combinations are shown in separate columns with rows corresponding to sample sizes 60, 90, 120. Shaded ribbons denote Monte-Carlo 95% confidence intervals based on 1,000 simulated trials per design point, and the dashed line indicates the nominal level $\alpha = 0.05$. Biomarker values are drawn from a Beta(0.5, 3) distribution indicating strong skewness.

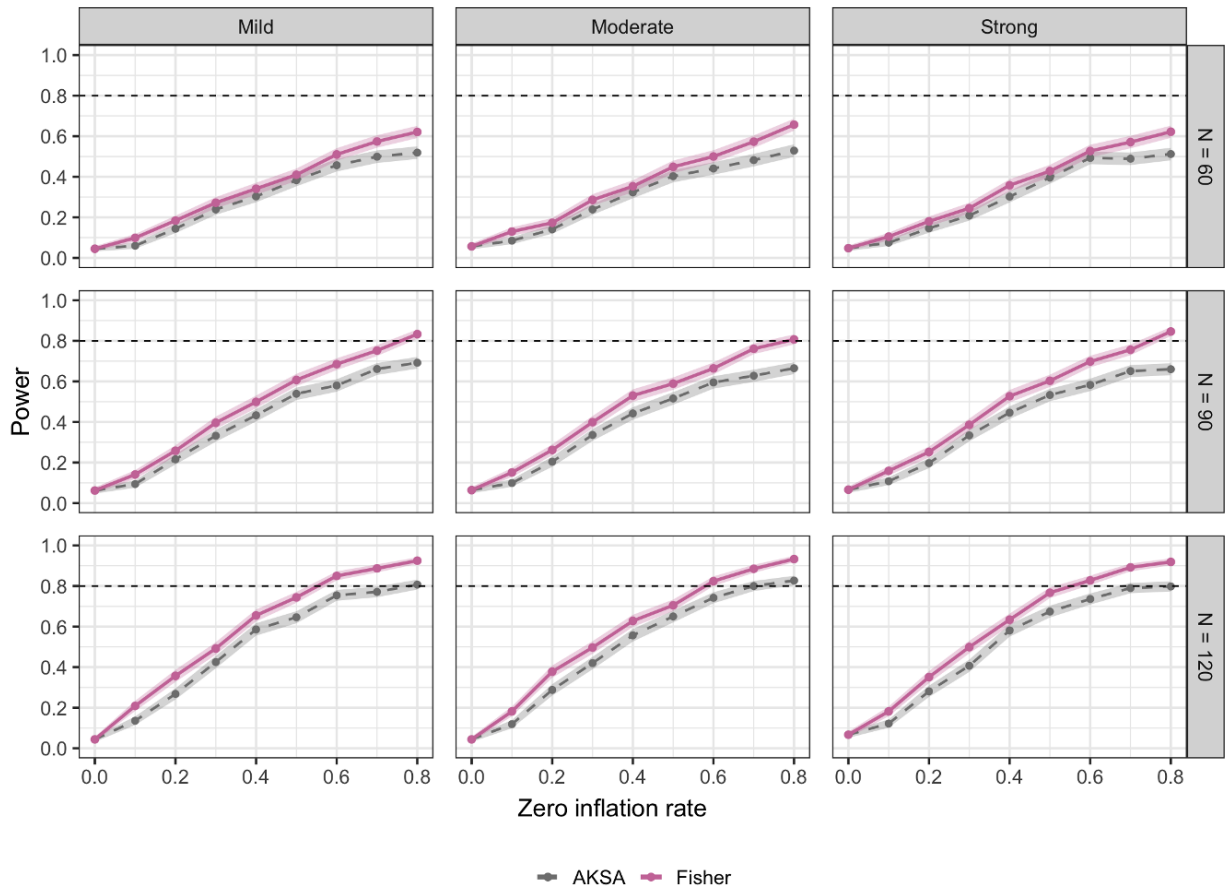

**Supplementary Figure S7. Empirical power of AKSA and Fisher's combination under the skewed biomarker and spike-only alternative.** Each panel displays the proportion of 1,000 simulated trials in which the corresponding test rejected the global null (y-axis) as the zero-inflation rate varied from 0 to 0.8 (x-axis). Columns distinguish the biomarker skewness, while rows represent sample size. Grey curves depict AKSA and magenta curves depict the two-step test combined by Fisher's method. Shaded ribbons are Monte-Carlo 95 % confidence intervals, and the horizontal dashed line marks the 80% power benchmark.

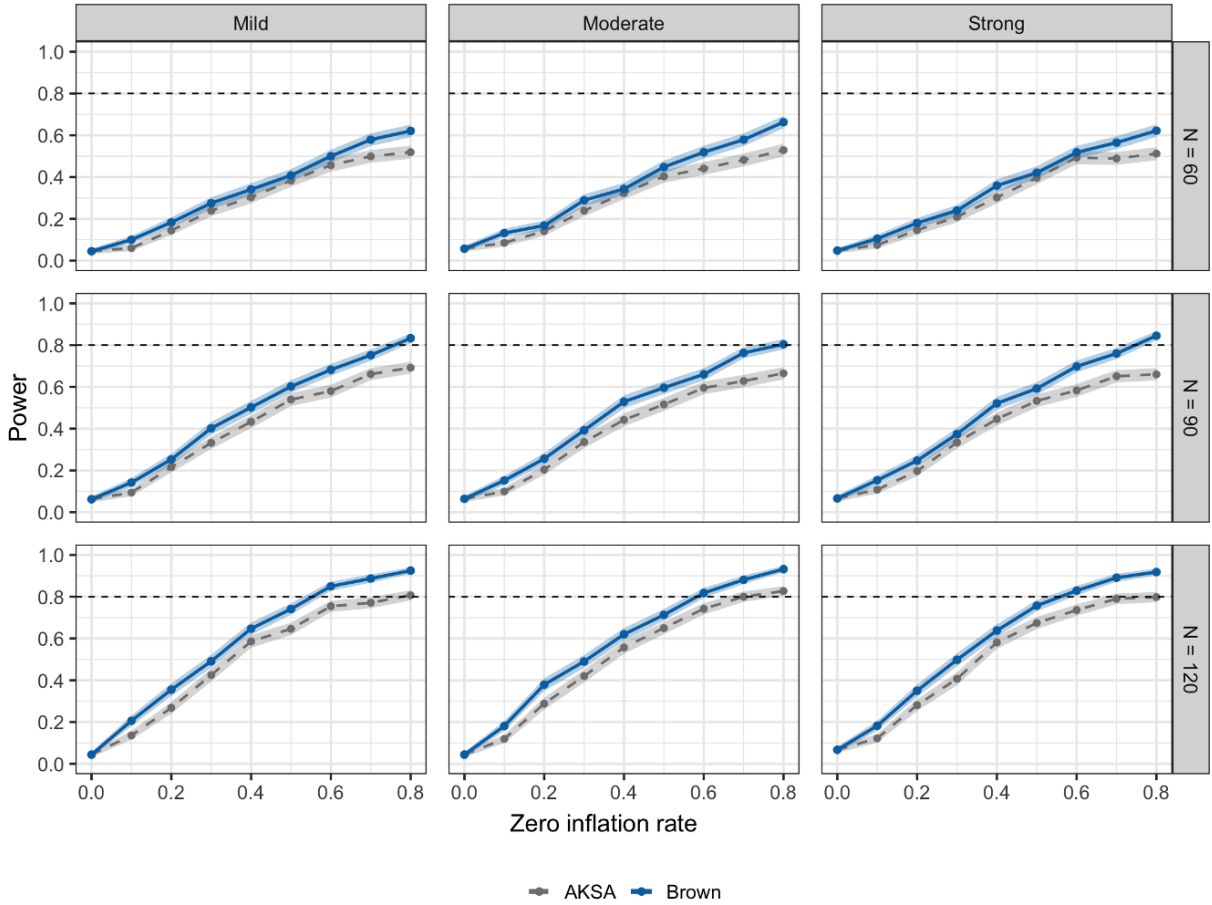

**Supplementary Figure S8. Empirical power of AKSA and Brown's combination under the skewed biomarker and spike-only alternative.** Each panel displays the proportion of 1,000 simulated trials in which the corresponding test rejected the global null (y-axis) as the zero-inflation rate varied from 0 to 0.8 (x-axis). Columns distinguish the biomarker skewness, while rows represent sample size. Grey curves depict AKSA and magenta curves depict the two-step test combined by Brown's method. Shaded ribbons are Monte-Carlo 95 % confidence intervals, and the horizontal dashed line marks the 80% power benchmark.

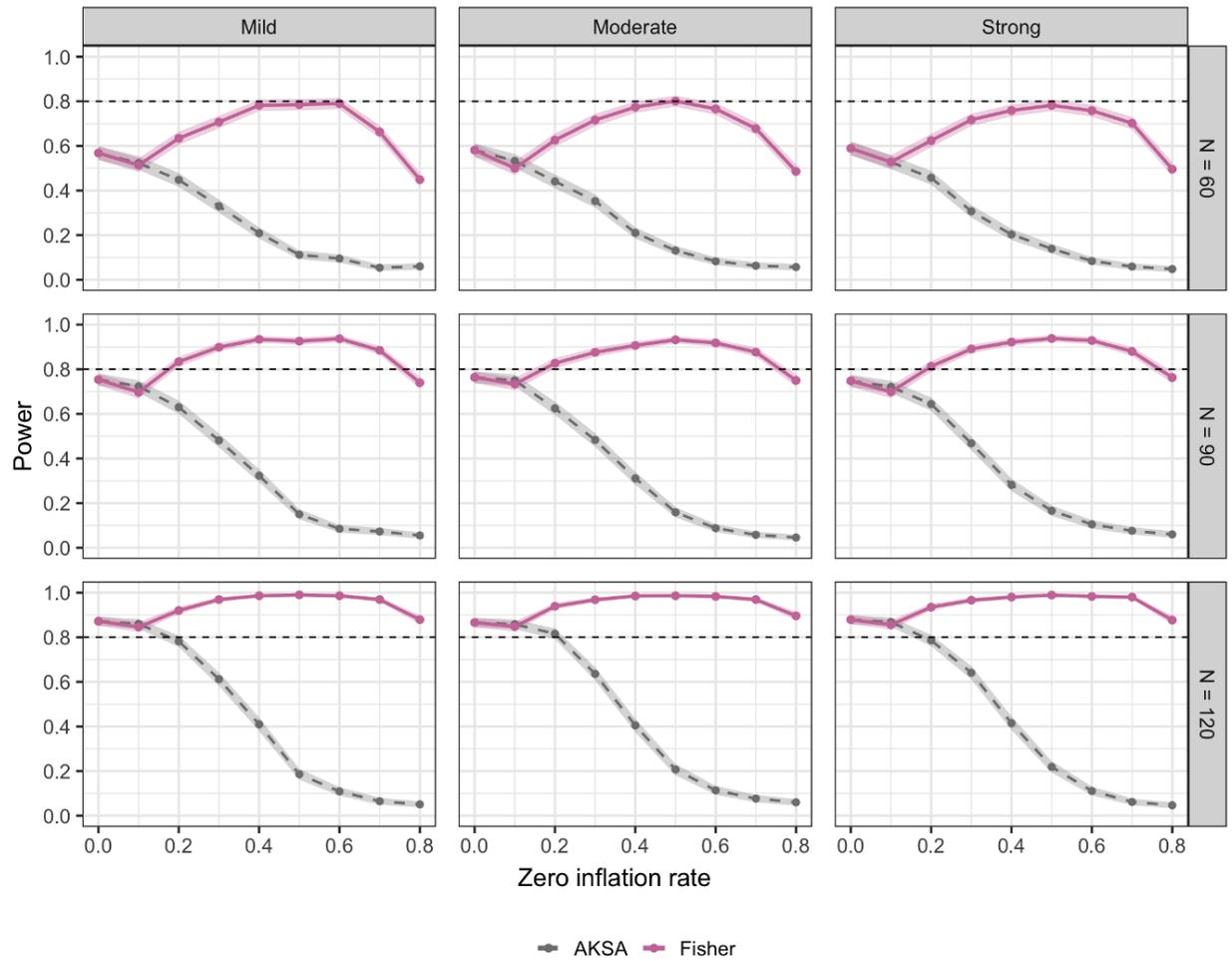

**Supplementary Figure S9. Empirical power of AKSA and Fisher's combination under the skewed biomarker and tail-only alternative.** Each panel displays the proportion of 1,000 simulated trials in which the corresponding test rejected the global null (y-axis) as the zero-inflation rate varied from 0 to 0.8 (x-axis). Columns distinguish the biomarker skewness, while rows represent sample size. Grey curves depict AKSA and magenta curves depict the two-step test combined by Fisher's method. Shaded ribbons are Monte-Carlo 95 % confidence intervals, and the horizontal dashed line marks the 80% power benchmark.

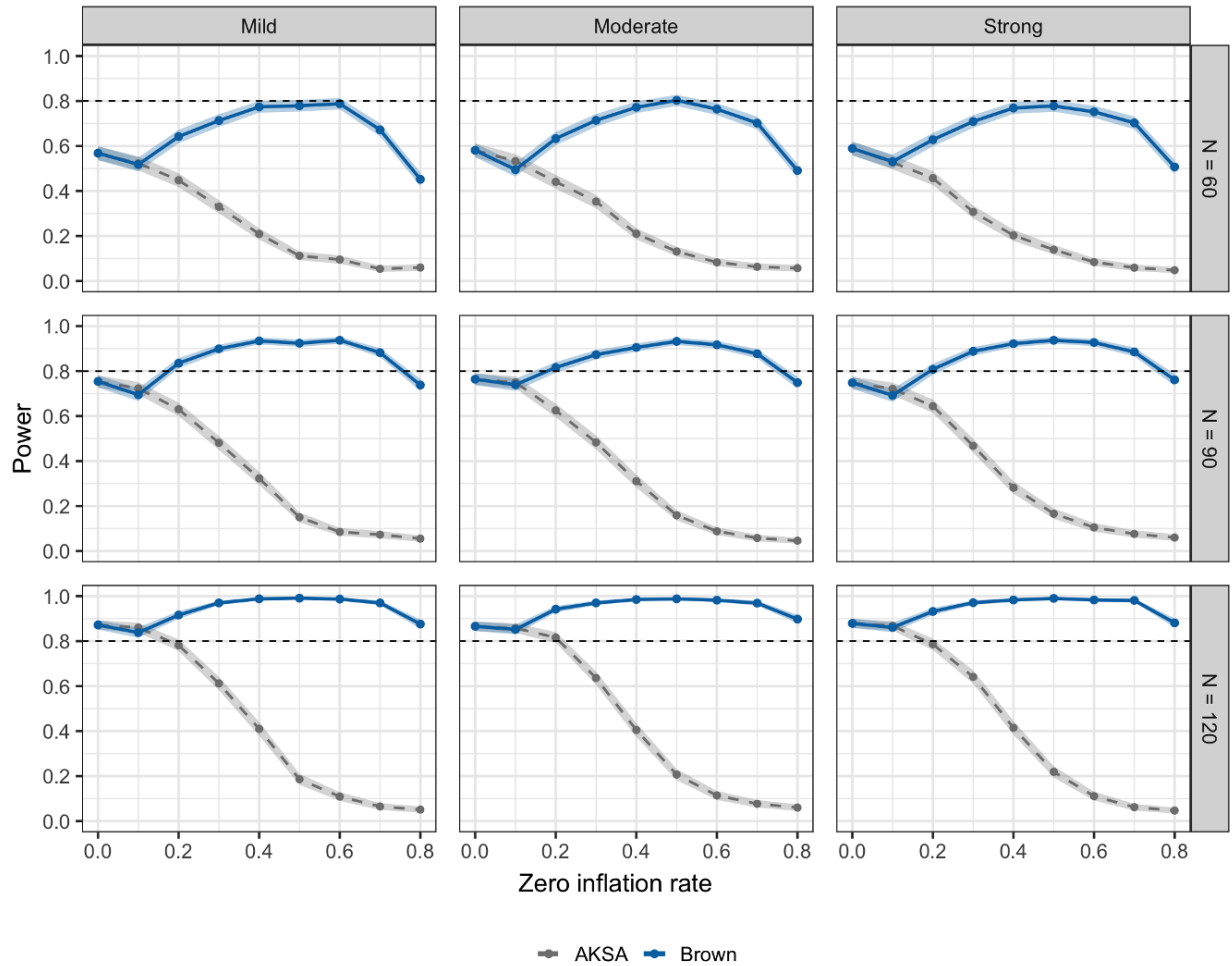

**Supplementary Figure S10. Empirical power of AKSA and Brown's combination under the skewed biomarker and tail-only alternative.** Each panel displays the proportion of 1,000 simulated trials in which the corresponding test rejected the global null (y-axis) as the zero-inflation rate varied from 0 to 0.8 (x-axis). Columns distinguish the biomarker skewness, while rows represent sample size. Grey curves depict AKSA and magenta curves depict the two-step test combined by Brown's method. Shaded ribbons are Monte-Carlo 95 % confidence intervals, and the horizontal dashed line marks the 80% power benchmark.

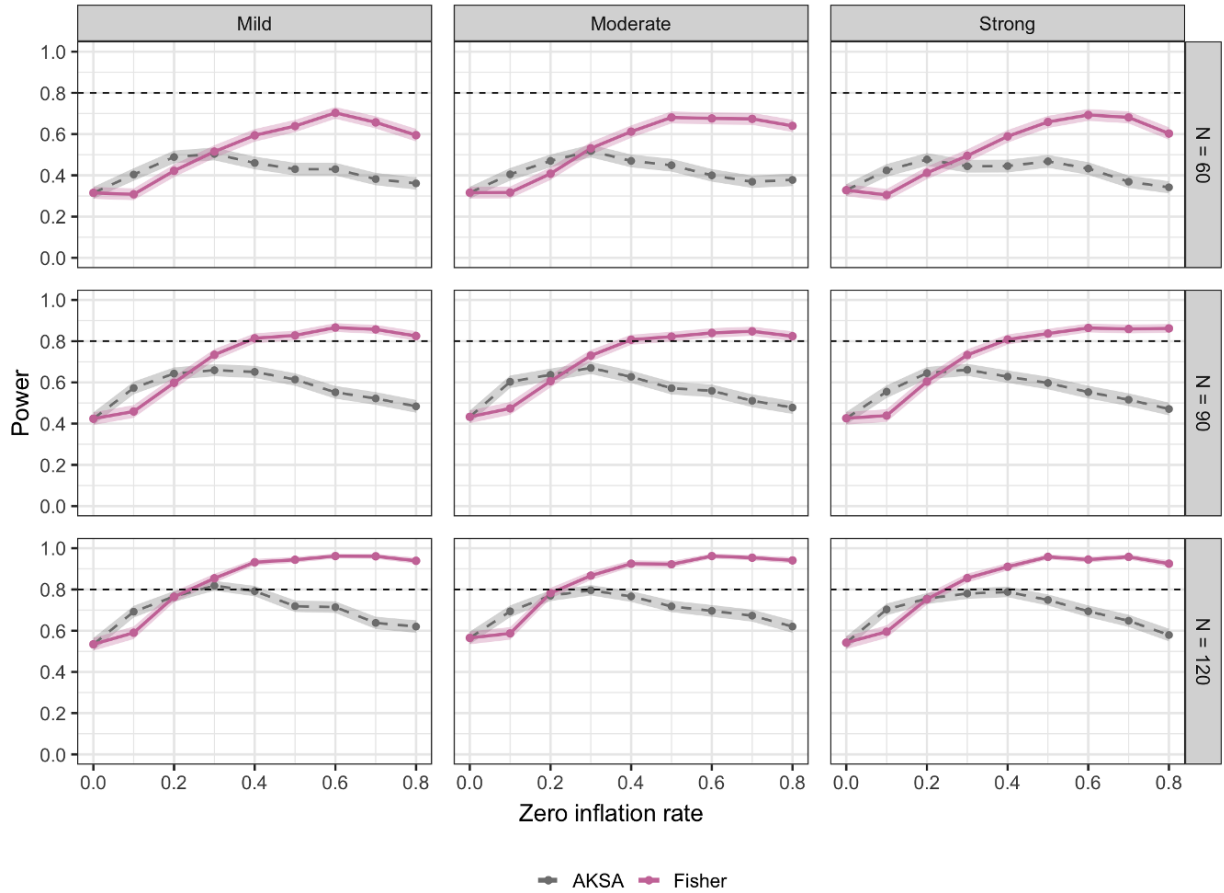

**Supplementary Figure S11. Empirical power of AKSA and Fisher's combination under the skewed biomarker and mix alternative.** Each panel displays the proportion of 1,000 simulated trials in which the corresponding test rejected the global null (y-axis) as the zero-inflation rate varied from 0 to 0.8 (x-axis). Columns distinguish the biomarker skewness, while rows represent sample size. Grey curves depict AKSA and magenta curves depict the two-step test combined by Fisher's method. Shaded ribbons are Monte-Carlo 95 % confidence intervals, and the horizontal dashed line marks the 80% power benchmark.

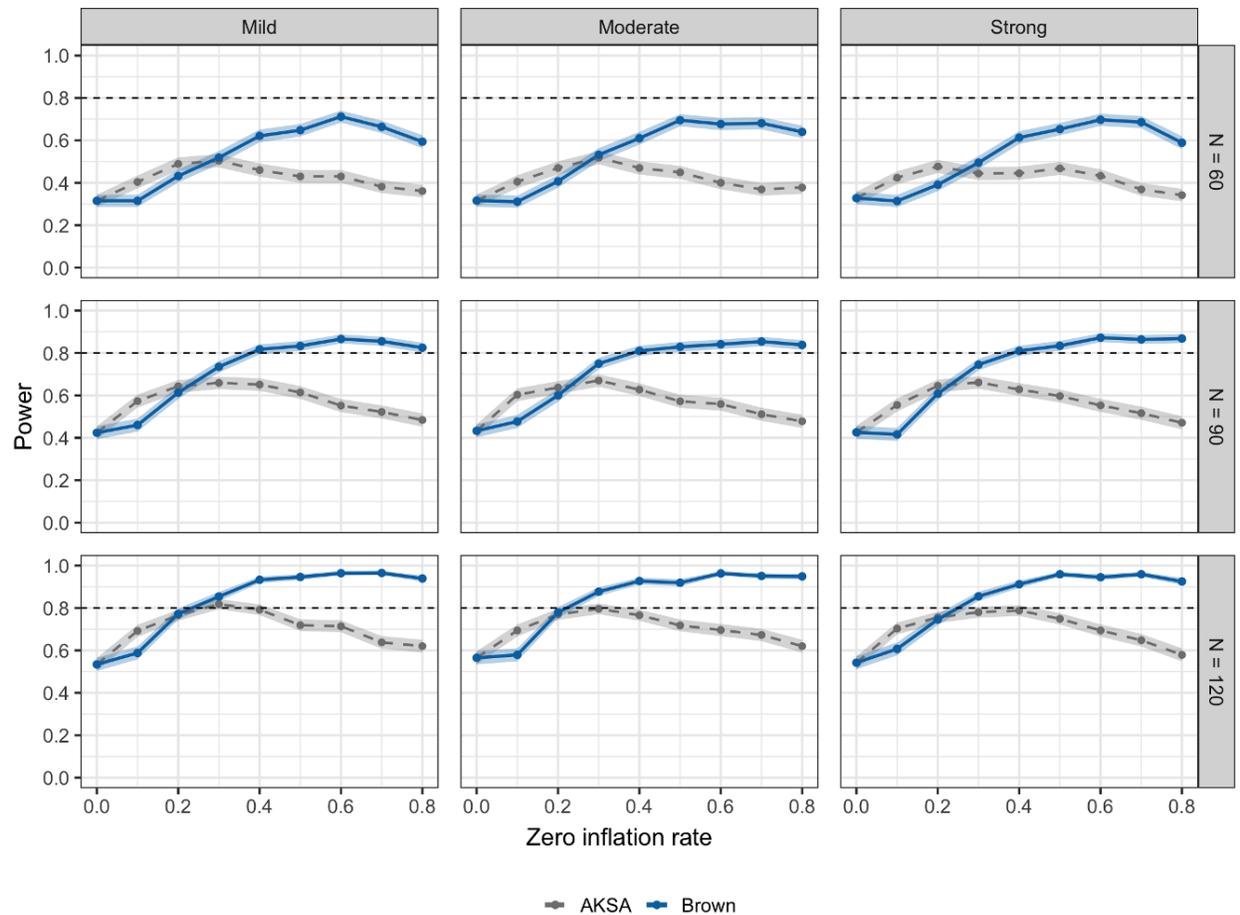

**Supplementary Figure S12. Empirical power of AKSA and Brown's combination under the skewed biomarker and mix alternative.** Each panel displays the proportion of 1,000 simulated trials in which the corresponding test rejected the global null (y-axis) as the zero-inflation rate varied from 0 to 0.8 (x-axis). Columns distinguish the biomarker skewness, while rows represent sample size. Grey curves depict AKSA and magenta curves depict the two-step test combined by Brown's method. Shaded ribbons are Monte-Carlo 95 % confidence intervals, and the horizontal dashed line marks the 80% power benchmark.